\documentclass[preprint,aps,prd,superscriptaddress,nofootinbib]{revtex4-1}

\usepackage{amsmath}
\usepackage{graphicx}
\usepackage{hyperref}
\usepackage[T1]{fontenc}
\usepackage{slashed}
\usepackage{bbold}
\usepackage{color}
\usepackage{url}
\usepackage{placeins}

\newcommand{\be}{\begin{equation}}
\newcommand{\ee}{\end{equation}}

\usepackage[normalem]{ulem}

\begin{document}

\title{Probing Light  Dark Matter with a Hadrophilic Scalar Mediator}

\preprint{PITT-PACC-1817}

\author{Brian Batell}
\email{batell@pitt.edu}
\affiliation{Pittsburgh Particle Physics, Astrophysics, and Cosmology Center,\\Department of Physics and Astronomy, University of Pittsburgh, Pittsburgh, USA}
\author{Ayres Freitas}
\email{afreitas@pitt.edu}
\affiliation{Pittsburgh Particle Physics, Astrophysics, and Cosmology Center,\\Department of Physics and Astronomy, University of Pittsburgh, Pittsburgh, USA}
\author{Ahmed Ismail}
\email{aismail@pitt.edu}
\affiliation{Pittsburgh Particle Physics, Astrophysics, and Cosmology Center,\\Department of Physics and Astronomy, University of Pittsburgh, Pittsburgh, USA}
\author{David McKeen}
\email{mckeen@triumf.ca}
\affiliation{TRIUMF, 4004 Wesbrook Mall, Vancouver, BC V6T 2A3, Canada}

\begin{abstract}
We investigate the thermal cosmology and terrestrial and astrophysical phenomenology of a sub-GeV hadrophilic dark sector. The specific construction explored in this work features a Dirac fermion dark matter candidate interacting with a light scalar mediator that dominantly couples to the up-quark. The correct freeze-out relic abundance may be achieved via dark matter annihilation directly to hadrons or through secluded annihilation to scalar mediators. A rich and distinctive phenomenology is present in this scenario, with probes arising from precision meson decays, proton beam dump experiments, colliders, direct detection experiments, supernovae, and nucleosynthesis. In the future, experiments such as NA62, REDTOP, SHiP, SBND, and NEWS-G will be able to explore a significant portion of the cosmologically motivated parameter space. 
\end{abstract}

\maketitle

\section{Introduction}
\label{sec:intro}
% !TEX root = up-DM.tex

The dark matter puzzle is among the most pressing problems in particle physics and cosmology today. In recent years, a growing experimental and observational program to search for non-gravitational dark matter interactions has coincided with a broader theoretical exploration of viable dark matter models, novel cosmological histories, and associated phenomenological signatures. Through these efforts, the paradigm of a light dark sector containing new (sub-)GeV-scale singlet dark particles coupled to ordinary matter through a light mediator has emerged as a compelling possibility and remains under active investigation. 
An excellent survey can be found in the recent community studies \cite{Battaglieri:2017aum,Alexander:2016aln}.

The experimental signatures of a particular dark sector model are, to a large extent, governed by the couplings of the mediator particle to ordinary matter. 
For instance, in the popular vector portal dark matter model~\cite{Boehm:2003hm,Pospelov:2007mp,Hooper:2008im,ArkaniHamed:2008qn}, the ``dark photon'' mediator couples to electrically charged particles via kinetic mixing with the ordinary photon~\cite{Galison:1983pa,Holdom:1985ag}. 
A rich phenomenology results due to the couplings to both leptons (e.g.,  electron beam fixed-target experiments, low-energy $e^+ e^-$ colliders, direct detection via electron scattering, lepton anomalous magnetic moments) and quarks (e.g., proton beam fixed-target experiments, direct detection via nuclear scattering, rare meson decays); see Refs.~\cite{Battaglieri:2017aum,Alexander:2016aln} and references therein. Besides kinetically mixed dark photons, a variety of mediators have been explored in the literature, including Higgs portal scalars~\cite{Krnjaic:2015mbs,Evans:2017kti}, neutrino portal fermions~\cite{Bertoni:2014mva,Gonzalez-Macias:2016vxy,Ibarra:2016fco,Escudero:2016ksa,Batell:2017rol,Batell:2017cmf,Schmaltz:2017oov}, and vector bosons coupled to anomaly free currents~\cite{Izaguirre:2014dua,Altmannshofer:2014cfa,Ilten:2018crw,Bauer:2018onh,Kahn:2018cqs,Berlin:2018bsc}, all of which have distinctive patterns of couplings to SM particles and resulting phenomenology.  Such wide theoretical investigation is required to identify the full range of phenomena associated with dark sectors and discern the physics potential of proposed searches and new experiments with respect to existing constraints over a broad range of models and couplings.  

It is in this context that we are motivated to explore dark sectors with hadrophilic  (or leptophobic) mediators. Finding viable constructions with light hadrophilic mediators is challenging in comparison to the mediators mentioned above. Leptophobic vector mediators coupled to, e.g., baryon number, have been considered in the past~\cite{Tulin:2014tya,Batell:2014yra,Soper:2014ska,Dobrescu:2014ita,Coloma:2015pih,Kouvaris:2016afs,Frugiuele:2017zvx,deNiverville:2018dbu} and are possible in principle. However, it was recently shown that such scenarios face stringent constraints due to enhanced production of the longitudinal mode in a variety of rare decays~\cite{Dror:2017ehi,Dror:2017nsg}, a result that can be traced to the anomalous nature of the baryon number symmetry. Leptophobic scalar mediators, on the other hand, face a different set of challenges, including new flavor-changing neutral currents (FCNC) and a naturalness problem related to the light scalar mass. Typically, a nontrivial flavor hypothesis (e.g., Minimal Flavor Violation~\cite{DAmbrosio:2002vsn} or alignment) on the scalar-quark couplings is needed to satisfy FCNC constraints, while naturalness arguments would point to weaker couplings for smaller scalar masses. An extensive analysis of these issues was recently carried out in Ref. \cite{Batell:2017kty}, where it was found that ``flavor-specific'' scalar couplings, i.e., scalars that couple to a specific quark mass eigenstate, can satisfy existing FCNC constraints even for relatively sizable couplings in the natural regions of parameter space. This framework therefore provides a promising point of departure to study a light leptophobic dark sector, and this is the task undertaken in this paper. We note that a similar analysis of the flavor-alignment hypothesis in BSM theories is presented in Ref.~\cite{Egana-Ugrinovic:2018znw}. In addition, previous work considered leptophobic scalars that couple dominantly to gluons or to top quarks in the sub-GeV range~\cite{Knapen:2017xzo}.  
Here we will instead focus on the case in which the scalar couples dominantly to the up-quark, finding some differences in comparison to the gluon-specific and top-specific scenarios that will be highlighted below. 

In particular, we find that a diverse and complementary set of experimental and observational probes, including precision meson decay measurements, beam dump experiments, colliders, direct detection, supernovae, and nucleosynthesis, already provides significant coverage of this scenario. Nevertheless, there are viable regions of parameter space where thermal freezeout can set the correct relic dark matter abundance, both through direct annihilation to hadrons as well as through secluded annihilation to scalar mediators.  
We also describe several promising ongoing or future experiments that will be able to further test this scenario, including NA62, REDTOP, SHiP, SBND, and NEWS-G.

The remainder of this paper is organized as follows. In the next section, we define the basic framework for the dark sector with up-quark specific scalar mediator couplings. In Section \ref{sec:pheno}, we investigate the phenomenological implications of this scenario for meson decays, beam dump experiments, colliders, direct detection, astrophysics, and cosmology. Section \ref{sec:concl} contains our conclusions. We have also included an Appendix which presents a description of the hadronic couplings for a general flavor-diagonal scalar mediator.

\section{Framework}
\label{sec:framework}
% !TEX root = up-dm.tex

The theoretical framework for a flavor-specific scalar mediator has been presented in Ref.~\cite{Batell:2017kty}, and we begin by reviewing its essential features. 
A singlet scalar mediator $S$ with mass $m_S$ is assumed to couple the visible and dark sectors. We will take dark matter to be a SM singlet Dirac fermion $\chi$ with mass $m_\chi$ that is charged under a $Z_2$ stabilizing symmetry. 
The mediator is assumed to couple dominantly to up quarks through a dimension-five operator generated at a UV scale $M$. 
The relevant terms in the Lagrangian are 
\be
\label{eq:bsmlag}
\begin{aligned}
\mathcal{L} \supset  &~  i \bar{\chi} (\slashed{D} - m_\chi) \chi + \frac{1}{2} \partial_\mu S \partial^\mu S - \frac{1}{2} m_S^2 S^2 
- \biggl(g_\chi  \, S \bar{\chi}_L  \chi_R +  \frac{c_S}{M} S \bar{Q}_L U_R H_c + \mathrm{h.c.} \biggr),\\
\rightarrow  &~  i \bar{\chi} (\slashed{D} - m_\chi) \chi + \frac{1}{2} \partial_\mu S \partial^\mu S - \frac{1}{2} m_S^2 S^2 
- \biggl(g_\chi  \, S \bar{\chi}_L  \chi_R +  g_u S \bar u_L u_R + \mathrm{h.c.} \biggr),
\end{aligned}
\ee
where in the second line we have defined the effective coupling 
\begin{equation}
g_u \equiv \frac{c_S v}{\sqrt{2}M},
\end{equation}
with $v = 246$ GeV the Higgs vacuum expectation value (vev). 
Other possible dimension-five operators involving $S$,$Q_L$,$U_R$, such as  $\partial_\mu S \bar{U}_R \gamma^\mu U_R$ and $\biggl( i S \bar{U}_R \slashed{D} U_R + \mathrm{h.c.} \biggr)$,
are related to the one written in Eq.~(\ref{eq:bsmlag}) by a field redefinition. 
Besides these, additional operators such as $S G^{\mu\nu} G_{\mu\nu}$ are expected on general grounds, but their presence and size depends on the particular UV completion. 
We will assume that the $c_S$ coupling in Eq.~(\ref{eq:bsmlag}) dominates over other dimension-five operators.

Under the $U(3)_Q \times U(3)_U \times U(3)_D$ global flavor symmetry the coupling $c_S$ in Eq.~(\ref{eq:bsmlag}), viewed as a spurion, transforms as $c_S \sim (3, \bar{3}, 1)$. The assumption that $S$ couples dominantly to the physical up quark,  such that in the mass basis $c_S \propto {\rm diag}(1,0,0)$, implies that $c_S$  breaks the flavor symmetry according to $U(3)_Q \times U(3)_U \rightarrow U(1)_u \times U(2)_{ctL} \times U(2)_{ctR}$, where the $U(1)_u$ factor is aligned with and identical to the one left unbroken by the up-quark Yukawa spurion. We emphasize that this symmetry breaking pattern is a hypothesis on the form of our low energy effective theory. It would certainly be worthwhile to explore UV model constructions which realize alignment and flavor-specific structures; for some promising directions along these lines, see  Refs.~\cite{Knapen:2015hia,Altmannshofer:2017uvs}.
Setting aside questions on the UV origin of such structures, the vast majority of dangerous FCNC processes are suppressed to acceptable levels once this flavor hypothesis is made. However, as we will see in Section~\ref{sec:pheno}, rare flavor changing meson decays (e.g., kaon decays) can still provide a sensitive probe if the scalars are light. 

We now turn to a discussion of the scalar potential for $S$. As is well-known, a light scalar with large couplings to heavy UV states is expected to receive large radiative contributions to its mass (quadratic) and, in the case of a singlet scalar, tadpole (linear) terms in the potential. A standard naturalness argument would then suggest that the lighter the scalar particle is, the smaller its couplings should be. This argument applies without any ambiguity to $c_S$ in Eq.~(\ref{eq:bsmlag}); since it is a dimension-five operator we must introduce degrees of freedom at the UV scale $M$, which would give a large correction to the scalar mass unless $c_S$ is sufficiently small. In addition, in the case of the singlet considered here, the tadpole induces a vev for $S$ and in turn a contribution to the effective up-quark Yukawa coupling. As discussed in detail in Ref.~\cite{Batell:2017kty}, the size of the leading two-loop corrections to the scalar mass and up-quark Yukawa (as well as other terms in the potential) are small provided
\be
\label{eq:natural-criterion}
\begin{aligned}
c_S \lesssim (16 \pi^2)\, \frac{m_S}{M} \approx (8 \times 10^{-3}) \left(\frac{m_S}{0.1 \, {\rm GeV}}\right) \left( \frac{2 \, {\rm TeV}}{M}\right), \\
\Longrightarrow ~ g_u \lesssim \frac{16 \pi^2}{\sqrt{2}}\, \frac{m_S v}{M^2} \approx (7 \times 10^{-4}) \left(\frac{m_S}{0.1 \, {\rm GeV}}\right) \left( \frac{2 \, {\rm TeV}}{M}\right)^2.
\end{aligned}
\ee
We note that the criterion in Eq.~(\ref{eq:natural-criterion}) is derived considering only the interactions and fields in the effective field theory (\ref{eq:bsmlag}). In particular, it often happens that in explicit UV completions there is a larger one-loop correction to the scalar mass which can result in a moderately more restrictive criterion than the one given in Eq.~(\ref{eq:natural-criterion}). In any case, it is worth emphasizing that one can of course explore regions of parameter space not satisfying (\ref{eq:natural-criterion}) at the expense of fine-tuning. For this work, we will use (\ref{eq:natural-criterion}) to provide a rough picture of the boundary between the natural and tuned regions of parameter space.

\subsection{Thermal Cosmology}

For $m_\chi < m_S$, and assuming there are no lighter states in the dark sector, $\chi$ will annihilate directly to SM particles through an $s$-channel mediator. 
In this case, there is a region in the $(m_\chi, g_\chi, g_u)$ parameter space that predicts the observed dark matter relic density assuming a standard thermal cosmology. 
We take this region as a motivated target in parameter space, 
though we note that the $\chi$ density may also be set by non-thermal mechanisms, such as an early $\chi$-$\bar{\chi}$ asymmetry. 
If $g_\chi$ has a non-zero phase, $s$-wave annihilation can occur, which is constrained for symmetric dark matter at the thermal level by Planck measurements of the Cosmic Microwave Background (CMB) for $m_\chi \lesssim 10$ GeV~\cite{Ade:2015xua} and by Fermi-LAT dwarf spheroidal galaxy observations for $m_\chi \lesssim 100~\mathrm{GeV}$~\cite{Ackermann:2015zua}. As we will be interested in $m_\chi \lesssim \mathcal{O}(\mathrm{GeV})$, we henceforth assume that $g_\chi$ is purely real, in which case the cross section for annihilation to SM final states is given by
\begin{equation}
(\sigma v_{\mathrm{rel}})_{\chi \bar{\chi} \to {\rm SM}} = \frac{g_\chi^2 m_\chi v_{\mathrm{rel}}^2 \Gamma_S\big|_{m_S = 2 m_\chi}}{2 \left( \left( m_S^2 - 4 m_\chi^2 \right)^2 + m_S^2 \Gamma_S^2 \right)}.
\label{eq:sigmav}
\end{equation}
For real $g_\chi$ we observe that the annihilation rate is velocity suppressed, and the CMB and gamma-ray constraints are trivially satisfied. 
We note that $\Gamma_S$ in Eq.~(\ref{eq:sigmav}) is the full $S$ width while $\Gamma_S\big|_{m_S = 2 m_\chi}$ is the width of the scalar evaluated at $m_S = 2 m_\chi$. For $m_S$ well above $\Lambda_{\mathrm{QCD}}$, we can use the perturbative width for $S \to u \bar{u}$, which is given by 
$\Gamma_{S\rightarrow u \bar u} = 3 g_u^2 m_S/ 8 \pi$. At lower scalar masses, we must take hadronic effects into account to calculate $\Gamma_S$; our procedure will be described below while the full details are given in the Appendix. Finally, for the thermally averaged annihilation cross section,  $\langle \sigma v_{\rm rel} \rangle$, we use $\langle v_{\rm rel}^2\rangle  = 6 /x$, with $x = m_\chi/T$ and $T$ the temperature. 

In the opposite regime, $m_\chi > m_S$, dark matter will dominantly annihilate to pairs of mediators (``secluded'' annihilation \cite{Pospelov:2007mp}) for most parameter choices. 
For purely real $g_\chi$, and in the limit $m_S \ll m_\chi$, the annihilation cross section is given by
\begin{equation}
(\sigma v_{\mathrm{rel}})_{\chi \bar{\chi} \to SS} = \frac{3g_\chi^4 v_{\rm rel}^2}{128 \pi m_\chi^2}.
\label{eq:sigmav-secluded}
\end{equation}
We note that this cross section is also velocity suppressed and thus safe from CMB and gamma-ray constraints. It also only depends on the dark matter - mediator coupling $g_\chi$, and the correct thermal abundance can be achieved provided the mediator has only a minuscule coupling $g_u$ with the SM to maintain kinetic equilibrium with the bath.

With real $g_\chi$ as motivated above, even a small imaginary component of $g_u$ will generate a large neutron EDM that is generically in tension with the experimental limit~\cite{Afach:2015sja}; see Refs.~\cite{Batell:2017kty,Seng:2014pba} for detailed discussions. We will therefore take $g_u$ to be real as well from here on so that the scalar interactions respect $CP$. 
While thermal DM annihilating through a real scalar interaction generally faces strong spin-independent direct detection bounds, we will be interested in the sub-GeV regime where these constraints are relatively weak. A full discussion of this issue will be postponed to Section~\ref{sec:pheno}.

\subsection{Hadronic couplings of $S$}

At scales between $\Lambda_{\mathrm{QCD}}$ and the cut-off scale $M$ of the dimension-five operator in Eq.~(\ref{eq:bsmlag}), the above Lagrangian provides a good description of the theory. Below the QCD scale, however, the interactions of $S$ are naturally written in terms of its hadronic couplings. Many of the probes of the sub-GeV dark sector involve sub-GeV momentum transfer, so it is important to establish such a description. This is analogous to studies carried out long ago for a light SM Higgs boson~\cite{Gunion:1989we,Voloshin:1985tc,Raby:1988qf,Grinstein:1988yu,Donoghue:1990xh,Truong:1989my} and more recent studies of light Higgs portal scalars (see, e.g.~\cite{Monin:2018lee,Winkler:2018qyg,McKeen:2008gd}). We will begin by discussing the use of the chiral Lagrangian to estimate the couplings of $S$ to hadrons at very low momentum transfer. At intermediate scales above several hundred MeV the breakdown of the momentum expansion and appearance of hadronic resonances renders the chiral Lagrangian description inappropriate, and we continue here by describing the standard form factors which parametrize the couplings between $S$ and the mesons at such energies. Here we briefly summarize the main results of both treatments for an up-specific scalar.  We have included an Appendix with a detailed analysis of these issues for a general flavor-diagonal scalar, which should find applications outside the scope of this work. 

First, by expanding the chiral Lagrangian in powers of momentum, we can obtain a low energy description of the interactions of $S$ with the pNGBs associated with chiral symmetry breaking in terms of the known meson masses. Full details of our treatment are contained in Appendix~\ref{app:ChiralL}. Our starting point is the effective Lagrangian
\be
{\cal L} \supset \frac{f^2}{4}{\rm tr} [ (D_\mu \Sigma)^\dag D^\mu \Sigma] +  \frac{f^2}{4} {\rm tr}\!\left[\Sigma^\dag \chi + \chi^\dag \Sigma \right], 
\label{eq:L-p2-S}
\ee
where $\Sigma = e^{2 i \pi / f}$ contains the pNGB fields, $\pi = \pi^a T^a$, $f \approx 93$~MeV is the pion decay constant, and the scalar couplings of the quarks are contained in
\be
\chi = 2 B \left(
\begin{array}{ccc}
m_u  + g_u\, S & 0 & 0 \\
0 & m_d & 0 \\
0 & 0  & m_s
  \end{array}
  \right).
\ee
The spurion $\chi$ contains both the usual quark masses as well as the effective coupling of $S$ to quarks. 
In the above, $B$ is a dimensionful parameter that can be determined by expanding the Lagrangian to obtain the physically observed meson masses, yielding $B \simeq m_\pi^2/(m_u + m_d) \approx 2.6$~GeV. The Lagrangian~(\ref{eq:L-p2-S}) contains the meson mass terms and trilinear interactions of $S$ with the mesons\footnote{An $S$ tadpole induced by chiral symmetry breaking is present in Eq.~(\ref{eq:L-p2-S}) but is negligible compared to the radiative $S$ tadpole induced by UV physics at the scale $M$; see the discussion around Equation~(\ref{eq:natural-criterion}).}. As expected, the $S\pi\pi$ couplings are all of order $g_u B$. We have also included the $\eta'$ meson in our description; see the Appendix for details. 

At higher energies, chiral perturbation theory is insufficient to describe the interactions between $S$ and the mesons. We parametrize the couplings by form factors~\cite{Donoghue:1990xh,Monin:2018lee,Winkler:2018qyg}, e.g. for the pions
\begin{eqnarray}
\langle \pi(p) \pi(p') | m_u \bar{u} u + m_d \bar{d} d | 0 \rangle &=& \Gamma_\pi(s), \nonumber \\
\langle \pi(p) \pi(p') | m_u \bar{u} u - m_d \bar{d} d | 0 \rangle &=& \Omega_\pi(s),
\label{eq:formfactors}
\end{eqnarray}
as a function of the momentum transfer squared $s = (p + p')^2$.
In practice, the form factor $\Gamma_\pi$ and its kaon analog $\Gamma_K$ are obtained from pion and kaon scattering, with resonant effects leading to a significant departure from the predictions of the chiral Lagrangian, even if higher-derivative terms are considered in the latter. There is an ${\cal O}(1)$ spread among the results in the literature for these form factors, using different formalisms and experimental data, and we choose to simply take the recent results of~\cite{Winkler:2018qyg}\footnote{The impact of this choice on the parameter constraints shown in the next section are relatively minor, since it affects only a small region of parameter space.}.
For the isospin-breaking $\Omega$ form factors in~(\ref{eq:formfactors}), we are not aware of any experimental determinations in the literature and simply use the prediction arising from the chiral Lagrangian approach above. Although beyond our current scope, it would be worthwhile to investigate to what extent these form factors and also those involving $\eta$ mesons can be extracted from existing hadronic data. Further details and discussion regarding the form factors involving pions and kaons are given in the Appendix.

\subsection{Decays of the mediator $S$}

\begin{figure}
\centering
\includegraphics[width=0.49\textwidth]{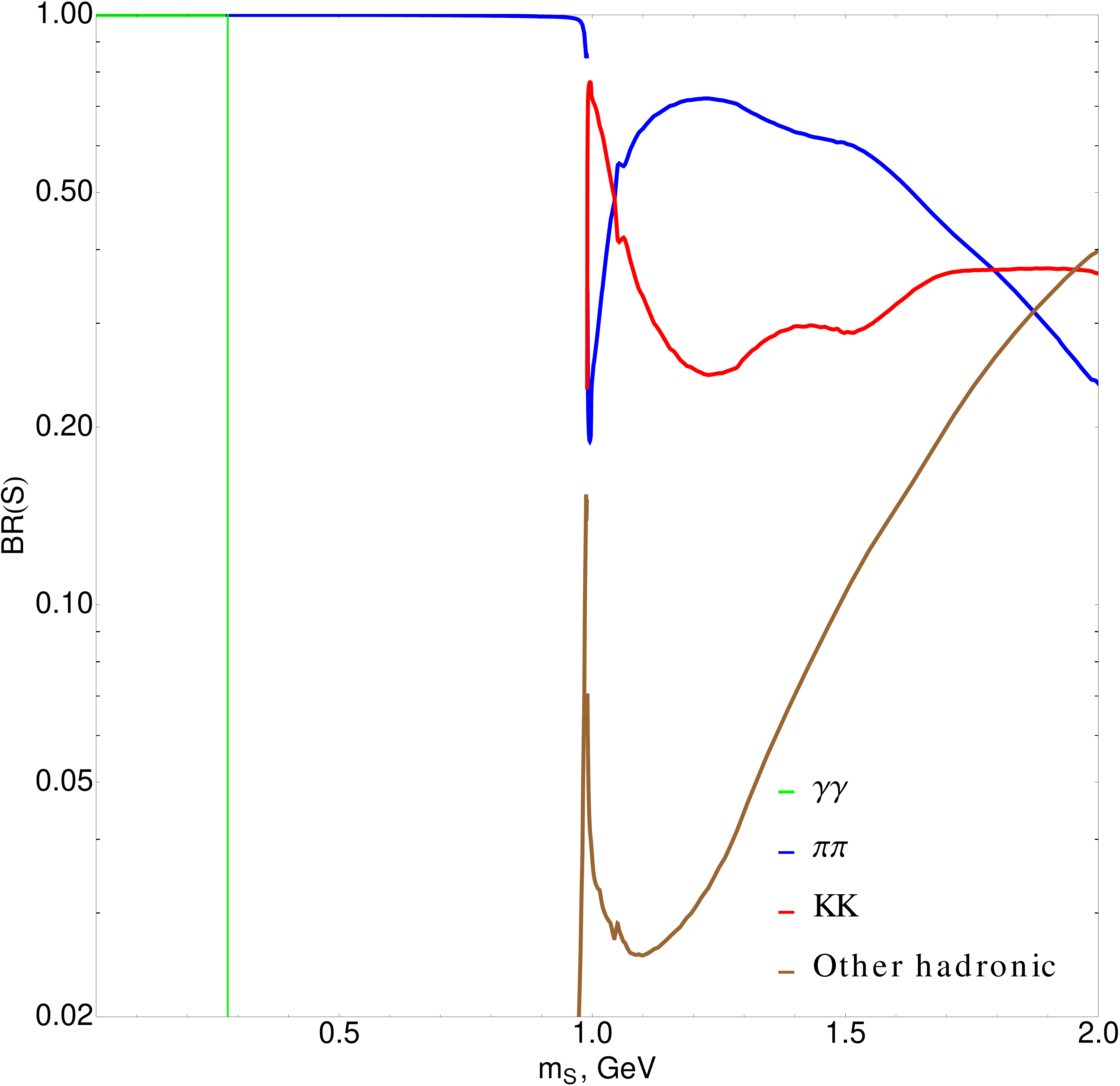}
\includegraphics[width=0.49\textwidth]{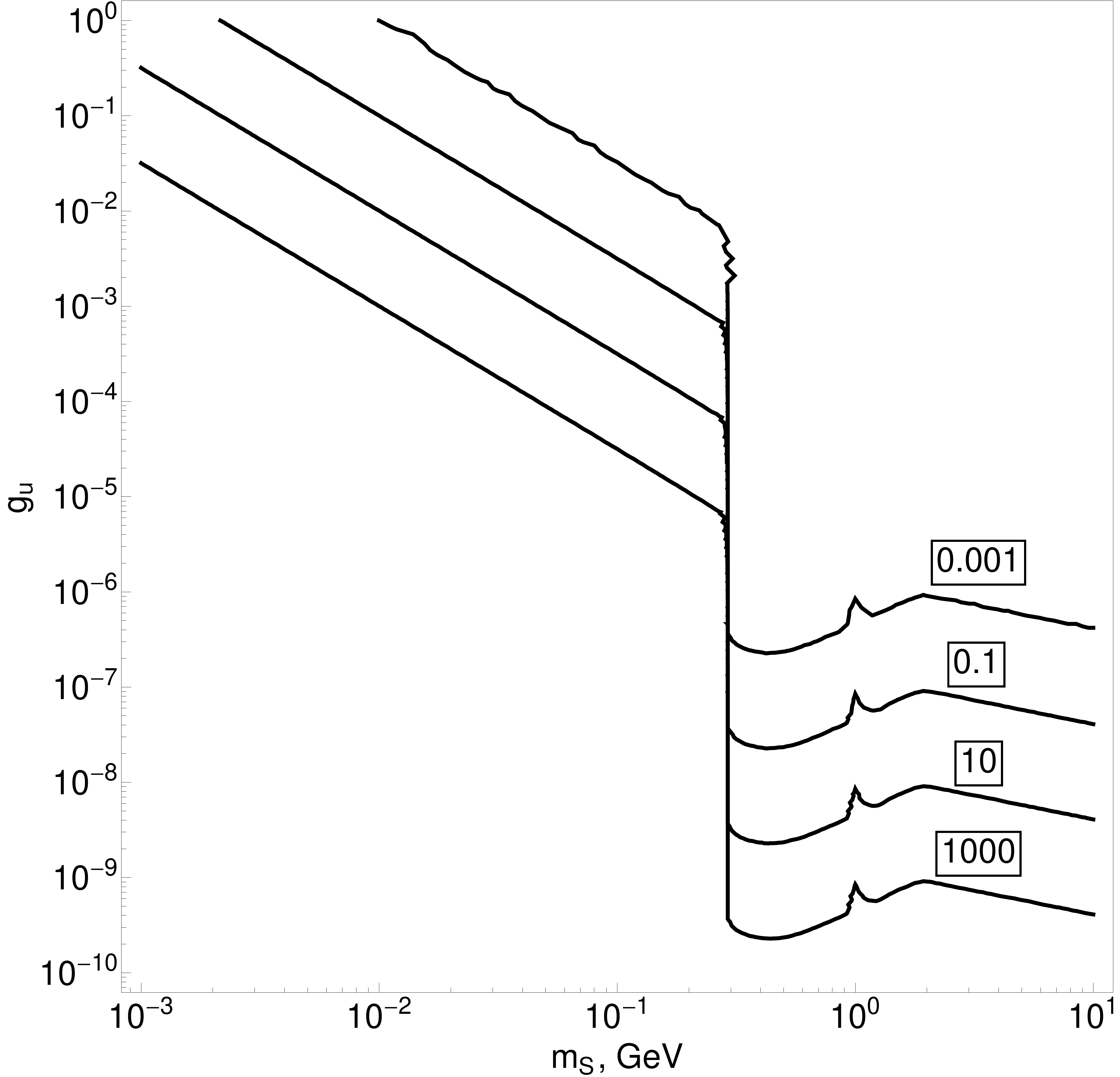}
\caption{\label{fig:brs} 
Left: Branching ratio of the scalar $S$ to SM final states, assuming that there are no dark sector decays of $S$. Right: Contours of constant decay length of $S$, in m, assuming only decays to SM final states.
}
\end{figure}

We are now ready to consider the decays of the mediator $S$, which are needed as input to the phenomenological analysis presented in Section~\ref{sec:pheno} as well as the dark matter annihilation cross section~(\ref{eq:sigmav}). For $m_\chi < m_S / 2$, the decay $S \to \chi \bar{\chi}$ occurs, and without a strong bound on $g_\chi$ we simply assume that it dominates the decays of $S$. On the other hand, for $m_\chi > m_S / 2$, $S$ decays exclusively to photons or hadrons, depending on its mass relative to the pion threshold $m_S = 2 m_\pi$.

For a scalar light enough to have no available tree level decay modes, the only available decay is to photons. The partial width for this decay is
\be
\Gamma_{S\to\gamma\gamma}=\sum_q \frac{\alpha^2 N_c^2 Q_q^4 g_q^2 m_S^3}{144\pi^3 m_q^2}\left| F_{1/2}\left(\frac{4m_q^2}{m_S^2}\right)\right|^2,
\ee
where the loop function is
\be
F_{1/2}\left(\tau\right)=\frac{3\tau}{2}\left[1+\left(1-\tau\right)\left(\sin^{-1}\frac{1}{\sqrt\tau}\right)^2\right]. \label{eq:f12}
\ee
The sum runs over all possible quarks in the loop, and when a first generation quark dominates, we use the constituent quark mass $\hat{m}_u = \hat{m}_d = 350~\mathrm{MeV}$.

For $m_S$ above the pion threshold, $S$ decays to hadrons. As for the case of DM annihilation, we must use the proper form factors for the interactions of $S$ with the mesons when $m_S$ is near $\Lambda_{\mathrm{QCD}}$. Again following~\cite{Winkler:2018qyg}, we match the partonic decay width for $S \to u \bar{u}$ at high $m_S$ to the sum of $S$ decays to mesons at low masses. 
Further details are contained in Appendix~\ref{app:ChiralL}. Fig.~\ref{fig:brs} shows the resulting branching ratios and decay lengths of the scalar. For $m_S < 2 m_\pi$, the only possible decay of $S$ is through a loop to photons. Above the pion threshold, $S \to \pi \pi$ is the only significant decay until kaons are kinematically accessible, strongly affecting $S$ decay through the $f_0(980)$ resonance. For $m_S \gtrsim 1.3~\mathrm{GeV}$, we start to lose predictive power on the individual hadronic decay channels of $S$ as the ``other hadronic'' decay channels, which we do not compute explicitly, are dominant. Nevertheless, we expect that our estimate for the $S$ total width is still reasonable. This calculation will have an impact in the next section, where the lifetime of $S$ is important in determining the relevant range of couplings $g_u$ that may be probed by a given experiment for fixed $m_S$.

\section{Phenomenology}
\label{sec:pheno}
% !TEX root = up-dm.tex

In this section, we explore the viable space for MeV-GeV scale dark matter with an up-philic scalar mediator. We describe current limits and future prospects from meson decays, beam dump experiments, Big Bang Nucleosynthesis (BBN), supernovae, direct detection, and colliders.

We will consider both the cases where the scalar decays visibly to SM particles, as described in Section~\ref{sec:framework}, as well as where the scalar decays invisibly to $\chi\bar{\chi}$. We first note that in the former case, there is a sharp change in the lifetime of $S$ at $m_S = 2 m_\pi$ as indicated by the right panel of Fig.~\ref{fig:brs}, due to the only available decay channel for lighter $S$ being the loop level decay to $\gamma \gamma$. For $m_S$ below the pion threshold, it is clearly possible that the scalar decays to SM states, but lives long enough that this decay is not observed in laboratory experiments, and so the scalar effectively appears to be invisible. Whether the photons from $S$ decay are visible in a given search depends on the detector setup as well as the energy with which the scalar is produced. Generally, these photons can be seen either if the coupling is large such that the loop decay proceeds rapidly, or if the scalar is produced with some boost and the detector is sufficiently far away from the point of production. Below, we will show examples of both types of searches which can directly observe the decay $S \to \gamma \gamma$. Otherwise, searches that would normally apply to an invisibly decaying $S$ can still be used. Here, the requirements are the converse: the coupling must be low enough or the detector must be sufficiently small that $S$ decays outside of it. Above the pion threshold, the decay of $S$ to pions generally proceeds promptly except at very low couplings, where $S$ can still be long-lived. When the decay is prompt, the $S$ can in principle be detected as a pion resonance. In practice, we will see that there is only a small region of $m_S$ where these decays provide a useful constraint on $g_u$.

We will also discuss the scenario where the new scalar decays invisibly, assuming that the branching fraction of $S \to \chi \chi$ is approximately 1, and furthermore that this decay occurs promptly. This is expected if $m_\chi < m_S / 2$ and the coupling $g_\chi$ is larger than the SM coupling of $S$. 
In the following, when considering a possible invisible decay mode of the scalar, we will assume that it dominates.

In principle the couplings $g_u, g_\chi$ and the masses $m_\chi, m_S$ are free parameters in our scenario. However, beyond the distinction of whether $S$ decays visibly or invisibly, the mass ratio $m_\chi / m_S$ has little practical impact on many searches for $S$ that involve only the $S\bar{u}u$ coupling, such as those at fixed-target experiments. As long as $g_\chi$ is sufficiently larger than $g_u$, changing it does not lead to significantly different phenomenology\footnote{In the region $2 m_\chi \approx m_S$, the relic density calculation is affected significantly by resonant annihilation.}.
 With this in mind, we will generally show limits in the $m_S$--$g_u$ plane. 
We note that for $m_\chi < m_S$ the thermal DM target and direct detection limits do not occupy a unique position on this plane, owing to the freedom to choose $m_\chi$ and $g_\chi$. On the other hand, for $m_\chi > m_S$, the thermal annihilation scenario is completely independent of the SM couplings of $S$ since secluded annihilation $\chi \bar{\chi} \to S S$ depletes the $\chi$ abundance; thus the correct relic density can be obtained anywhere in the $m_S$--$g_u$ plane, with an implied relation between $m_\chi$ and $g_\chi$ from Eq.~(\ref{eq:sigmav-secluded}). 
With these points in mind, we now explore the phenomenology of our scenario, with an eye towards both differences from the usual Higgs-like scalar case and opportunities for improvement in the future.

\subsection{Meson decays}

For sub-GeV $m_S$, there are many possible meson decays that contain $S$ in the final state, and through these decays $S$ can be copiously produced at precision decay experiments.
Unlike a Higgs-like scalar, for which meson-induced scalar production typically involves the top quark coupling in a penguin diagram, a scalar which couples predominantly to the first generation is produced directly from the quark content of the mesons. 
We estimate the production of scalars from the decays of $\eta$ and $K$ mesons using the chiral Lagrangian of Eq.~(\ref{eq:L-p2-S}).

The resulting branching ratios for the leading decays of the mesons which produce $S$ are
\begin{eqnarray}
{\rm Br}(\eta \rightarrow \pi^0 S) & = & \frac{c_{S\pi^0\eta}^2 g_u^2 B^2}{16 \pi m_\eta \Gamma_\eta} \lambda^{1/2}\left(1, \frac{m_S^2}{m_\eta^2},\frac{m_{\pi^0}^2}{m_\eta^2}\right)
\simeq 0.056 \left( \frac{g_u}{7 \times 10^{-4}} \right)^2, \label{eq:Breta-Pi-S}  \\
{\rm Br}(\eta' \rightarrow \pi^0 S) & = & \frac{c_{S\pi^0\eta'}^2 g_u^2 B^2}{16 \pi m_{\eta'} \Gamma_{\eta'}} \lambda^{1/2}\left(1, \frac{m_S^2}{m_{\eta'}^2},\frac{m_{\pi^0}^2}{m_{\eta'}^2}\right)
\simeq 1.3\times 10^{-4} \left( \frac{g_u}{7 \times 10^{-4}} \right)^2, \nonumber \\
{\rm Br}(K^+ \rightarrow \pi^+ S) &=& \frac{g_u^2 G_F^2 f_\pi^2 f_K^2 B^2}{8 \pi m_{K^+} \Gamma_{K^+}} |V_{ud} V_{us}|^2 \lambda^{1/2}\left(1, \frac{m_S^2}{m_{K^+}^2},\frac{m_{\pi^+}^2}{m_{K^+}^2}\right) \label{eq:BrK-Pi-S}  \simeq 3.2 \times 10^{-6} \left(\frac{g_u}{ 7 \times 10^{-4}}\right)^2, \nonumber
\end{eqnarray}
where $\lambda(a,b,c)= a^2+b^2+c^2 -2ab-2ac-2 bc$, $B \simeq m_\pi^2/(m_u+m_d) \approx 2.6$ GeV, and the coefficients $c_{S\pi^0\eta(')} = \frac{1}{\sqrt{3}} \cos \theta \mp \sqrt{\frac{2}{3}} \sin \theta$ parametrize the effects of $\eta-\eta'$ mixing, with $\theta \approx -20^\circ$. In the final expression of each line, we have taken $g_u$ close to its maximal ``natural'' value given in Eq.~(\ref{eq:natural-criterion}) under the assumption of a 100 MeV scalar. Eq.~(\ref{eq:Breta-Pi-S}) reveals the potential for significant production of scalars $S$ in the decays of light-flavor mesons. 

We briefly comment on the production of $S$ from heavy meson decays, e.g., $B \rightarrow K S$, where the chiral Lagrangian approach is not applicable. This is typically one of the largest production channels for a Higgs-like scalar, but turns out to be irrelevant for an up-philic scalar. 
We have estimated two possible contributions to this channel. First, we have computed the tree level weak decay with emission of $S$ from the spectator up-quark, dressed with meson form factors and wave functions parametrizing the momentum distribution of the quarks within the mesons~\cite{Aditya:2012ay}. However, due to the significant $V_{ub}$ suppression  in the amplitude, we find a branching ratio that is 1-2 orders of magnitude below current limits~\cite{Grygier:2017tzo} even for order one $g_u$. Second, we note that in principle $S$ has an induced interaction with gluons at two loops, which in turn feeds into an $S \bar{t} t$ coupling, giving penguin diagram contributions to the decay. However, the combination of the $S$ shift and up quark chiral symmetries implies that any $S G^{\mu \nu} G_{\mu \nu}$ interaction be proportional to both $g_u$ and $Y_u$. Together with the loop suppression, the resulting contribution to $B$ decays to the scalar is far too small to be of any importance. This behavior is in contrast to scalars whose interactions are dominated by top quark and gluon couplings, where searches for rare heavy meson decays set relevant bounds~\cite{Knapen:2017xzo}. Given these estimates, we do not consider limits from $B$ decays further.

We now turn to specific experiments which can exploit the meson decays of Eq.~(\ref{eq:Breta-Pi-S}). In many cases, there are direct measurements of exclusive meson decay channels. First, the decay $\eta \to \pi \gamma \gamma$ has been measured at the Mainz Microtron (MAMI) with 10\% precision~\cite{Nefkens:2014zlt}. We use this measurement to constrain the decay $\eta \to \pi S, S \to \gamma \gamma$~\cite{Liu:2018qgl}, conservatively requiring only that the branching ratio of such decays which occur within the Crystal Ball detector be smaller than the $2 \sigma$ upper bound on ${\rm BR}(\eta \to \pi \gamma \gamma)$ of $3.0 \times 10^{-4}$, i.e., making no assumption on the SM contribution to this final state. To calculate the fraction of the decays which produce photons within the detector, we use the fact that the $\eta$ is produced nearly at rest within the detector of radius 25 cm~\cite{McNicoll:2010qk}, which in our scalar mass region of interest puts a lower bound on $g_u$.

In addition, $4.7\times10^6$ $\eta \to \pi^0\pi^+\pi^-$ decays have been recorded by KLOE~\cite{Anastasi:2016cdz}, with a Dalitz analysis that shows no signs of resonances as would be produced if there were a contribution from $\eta \to \pi^0 S, S \to \pi^+ \pi^-$. For a given value of $m_S$, we estimate the background in the corresponding $m_S=m_{\pi^+\pi^-}$ bin by fitting the data in the other bins using a quadratic polynomial in $m_{\pi^+\pi^-}$ to describe the SM $\eta\to3\pi$ matrix element. We then require that the background plus the contribution from $\eta\to\pi^0 S$ not exceed the $3\sigma$ upper limit in this bin. For the values of $g_u$ probed here, the $S$ is very narrow and does not populate multiple $m_{\pi^+\pi^-}$ bins. This limit is valid in the mass range $2 m_\pi < m_S < m_\eta - m_\pi$. The proposed REDTOP experiment~\cite{Gatto:2016rae} hopes to produce around $10^{13}$~$\eta$/year using a $1.9~\rm GeV$ $p$ beam on a beryllium target. Given the branching ratio for $\eta \to \pi^0\pi^+\pi^-$ of $23\%$ this could result in a sample of $\pi^0\pi^+\pi^-$ events about $10^5$ times larger than at KLOE, which could allow the limit on $g_u$ to be improved by a factor of $\sim 20$.

For an invisibly decaying scalar, the analogous $\eta$ decay channel $\eta \to \pi S, S \to \chi \bar{\chi}$ could be considered. However, there is currently no search for $\eta \to \pi +~\mathrm{invisible}$. Performing such a search at the upcoming REDTOP experiment would be challenging since the signature would be simply two photons (that reconstruct a $\pi^0$) recoiling against missing momentum. However, because the longitudinal momentum of the $S$ is unknown, reconstructing the $\eta$ would be challenging. A second phase at REDTOP is also envisioned, using a higher energy beam to produce around $10^{11}$~$\eta^\prime$/year. The branching for $\eta^\prime\to\pi^0\pi^0\eta$ is $23\%$ and the extra kinematical constraints from the $\pi^0\to\gamma\gamma$ decays could allow the ``tagging'' of an $\eta$ that decays partially invisibly. A branching of ${\rm Br}(\eta \rightarrow \pi^0 S)\simeq 10^{-7}$, corresponding to $g_u\simeq10^{-6}$, would lead to ${\cal O}(10^3)$ such events. A detailed study of the prospects at REDTOP for this decay mode would be highly desirable.

Precision kaon measurements can also constrain our scalar, through the decay $K \to \pi S$. In particular, experiments E787 and E949 at Brookhaven searched~\cite{Adler:2002hy,Adler:2004hp,Adler:2008zza,Artamonov:2009sz} for the decay $K \to \pi \nu \bar{\nu}$, which constrains the rare decay $K \to \pi S$ where $S$ is either long-lived or decays invisibly. When $S$ decays visibly but $m_S < 2 m_\pi$ and $g_u$ is sufficiently small, $S$ does not decay within the E949 detector, and we calculate the expected number of scalars from $K \to \pi S$ which would produce such ``invisible'' signatures by asking how many scalars would decay to photons outside the 1.45 m radius of the E949 detector, as a function of $g_u$ and $m_S$. The upper edge of the exclusion region in the $m_S$ -- $g_u$ plane in the left panel of Fig.~\ref{fig:suulimits} corresponds to the coupling where the $S$ decays inside the detector, so that the scalar is no longer effectively invisible. For a genuinely invisibly decaying scalar, of course, the same search applies without an upper limit on $g_u$.

The $K \to \pi \nu \bar{\nu}$ mode will be measured to 10\% precision by the NA62 experiment~\cite{NA62:2017rwk}, which uses the 400 GeV CERN SPS proton beam to produce a secondary 75 GeV kaon beam, allowing the study of order $10^{13}$ decaying kaons over the life of the experiment. To estimate the increased sensitivity, we scale the E787/E949 result by the increase in the precision on the SM branching ratio for the decay being studied, following similar projections for axion-like particles~\cite{Izaguirre:2016dfi}. The actual eventual reach of NA62 will also depend on the theoretical uncertainty in the SM prediction for the decay branching ratio, which is currently 10\%~\cite{Buras:2015qea}; we neglect this here.

Lastly, the decay $\eta' \to \pi S$ can be used to constrain even heavier scalars than those probed by $\eta$ or kaon decays. Due to the strength of the other meson bounds, we focus only on the mass region $m_K - m_\pi < m_S < m_{\eta'} - m_\pi$, above the kinematic reach of kaon decays. If it decays to SM particles, a scalar of this mass would contribute to the decay $\eta' \to 3 \pi$, which has been measured by BESIII~\cite{Ablikim:2016frj}. We require that for such a scalar, the branching ratio for $\eta' \to \pi S$ not exceed the measured branching ratio for $\eta' \to 3 \pi$. On the other hand, if $S$ decays invisibly, we impose that the partial width $\Gamma(\eta' \to \pi S)$ not exceed the PDG fit for the total $\eta'$ width~\cite{Tanabashi:2018oca} of 196~keV. Additionally, a high energy REDTOP run could increase the number of reconstructed $\eta^\prime$ substantially.

\subsection{Beam dumps}

Having discussed precision decay measurements, we now describe the impact of beam dump experiments that can search for $S$. At proton beam dumps, because of the significant production cross section for $\eta$ mesons and its narrow width, $\eta \to \pi S$ can provide a significant scalar production channel. Furthermore, the $\eta'$ production rate at proton beam dumps is typically less than that of $\eta$ by about an order of magnitude, but allows for additional access to the scalar mass region $m_\eta - m_\pi < m_S < m_{\eta'} - m_\pi$; note that as this region is above the pion threshold, the additional gain is at very low coupling. 
Although $m_K$ and $m_\eta$ are similar, the branching fraction of kaons to scalars is smaller than that of $\eta$ to scalars at the same coupling, and kaons are produced less copiously in proton beam dumps. Therefore, kaon-induced production of $S$ is comparatively small, and we neglect it below. We emphasize again the difference with the case of a Higgs-like scalar: there, $B$ and $K$ mesons are the usual meson decay sources of production at beam dumps. However, for a scalar that preferentially couples to the first generation, the effective coupling of the heavy mesons to the scalar is very small, as discussed above, and the primary sources of scalars from meson decays are instead the light mesons. At proton fixed-target experiments, hadrophilic scalars can also be produced through bremsstrahlung, but we consider only meson decays here. It would be useful in the future to perform a detailed calculation of the additional bremsstrahlung production mode, which would improve the limits shown here.

The CHARM collaboration performed a search for axion-like particles produced from the 400 GeV SPS proton beam, decaying to $\gamma \gamma$ or leptons producing electromagnetic showers, in a detector located 480 m downstream and approximately 10 mrad off axis~\cite{Bergsma:1985qz}. We recast this search to place a limit on scalars produced from $\eta$ decays. To obtain the total $S$ production rate, we use CHARM's estimate of pion production, and then apply a scaling factor obtained from a simulation in CRMC~\cite{crmc} showing that approximately one $\eta$ is produced for every ten neutral pions within the geometrical acceptance of the CHARM detector~\footnote{Approximately one $\eta'$ is produced for every hundred neutral pions in the CHARM detector, and we include the effect but it does not yield any additional exclusion power.}. The distribution of the $\eta$ energy affects the boost of the scalar produced in $\eta \to \pi S$, and we simply assume that all $S$ particles are produced with energy 25 GeV, the average of the $E_\eta$ spectrum~\cite{Bergsma:1985qz}, as in previous studies~\cite{Alekhin:2015byh}. Combined with ${\rm Br}(\eta \rightarrow \pi^0 S)$ above and the lifetime estimates for $S$ from the previous section, we may then calculate the number of $\gamma \gamma$ events that would have been seen by CHARM for $m_S < 2 m_\pi$. As no events were seen, we simply require that no more than three $\gamma \gamma$ events would have occurred in the CHARM detector, assuming perfect reconstruction efficiency. Above the pion threshold, we note that each decay $S \to \pi^0 \pi^0$ would produce $4\gamma$ events, which would have again been visible at CHARM as electromagnetic showers. We can thus still place a  bound from the CHARM search for $m_\pi > 2 m_S$, occupying a region at much lower coupling due to the tree level decay of the scalar.

Future proton beam dumps will also be sensitive to $S$ production from mesons. We consider the case of SHiP~\cite{Alekhin:2015byh} in complete analogy with CHARM. SHiP would again use the 400 GeV SPS proton beam to produce very weakly coupled light particles, though with a much closer detector than CHARM, only about 70 m from the interaction point. Again we use CRMC to simulate $\eta(')$ production from proton interactions with the target material, and employ the same assumptions on the scalar energy spectrum and reconstruction efficiency as for CHARM above. Unlike for CHARM, the $\eta'$ production is sufficient to probe an additional part of parameter space in the relevant mass regime at $g_u \sim 10^{-8}$. It would also be interesting to examine the sensitivity of other proposed detectors at the LHC and beyond targeting long-lived particles, including CODEX-b~\cite{Gligorov:2017nwh}, FASER~\cite{Ariga:2018uku}, MATHUSLA~\cite{Curtin:2018mvb}, and SeaQuest~\cite{Gardner:2015wea,Berlin:2018pwi}.

While thus far we have only discussed proton beam dumps, the loop-level coupling of our scalar to photons implies that $S$ may be produced through Primakoff production at electron beam dumps as well, analogously to the muon-philic case~\cite{Batell:2017kty}.
We have checked the limits on the loop-level interaction from searches for axion-like particles at experiment E137 at SLAC~\cite{Bjorken:1988as,Dobrich:2015jyk,Dolan:2017osp}. The resulting bound largely overlaps with that from CHARM, and does not constrain any additional parameter space once the other limits in this section are taken into account.

Now, when $S$ decays to dark matter, the standard beam dump tests searching for the decay products of new particles that are produced in fixed-target collisions and decay far downstream are no longer applicable. However, through the decay of $S$,
light DM can be produced in the primary collisions of protons in a beam dump and subsequently detected through its scattering with nucleons in a downstream detector~\cite{Batell:2009di,deNiverville:2011it,deNiverville:2012ij,Batell:2014yra,Dobrescu:2014ita,Kahn:2014sra,deNiverville:2015mwa,Coloma:2015pih,deNiverville:2016rqh,Izaguirre:2017bqb,Frugiuele:2017zvx,deNiverville:2018dbu}. 
This approach has been employed recently by the MiniBooNE-DM collaboration~\cite{Aguilar-Arevalo:2017mqx,Aguilar-Arevalo:2018wea,Dharmapalan:2012xp}, and the null result from their search in the nucleon elastic scattering channel~\cite{Aguilar-Arevalo:2017mqx} leads to relevant constraints on our scenario. 
To estimate the potential dark matter scattering yield and derive the constraints implied by the search~\cite{Aguilar-Arevalo:2017mqx} we have performed a Monte Carlo simulation using methods similar to those employed in the BdNMC package~\cite{deNiverville:2016rqh}. During the dedicated MiniBooNE-DM run, $1.86 \times 10^{20}$ protons-on-target (POT) from the Fermilab Booster (8 GeV kinetic energy) were directed onto an iron beam dump. Focusing on the regime $m_S > 2m_\chi$, we consider the following DM production chain: 1) Meson production in the primary collisions, $p N \rightarrow \eta(') + X$, followed by 2) meson decay to scalar mediator, $\eta(') \rightarrow \pi^0 S$,  followed by 3) mediator decay to dark matter, $S\rightarrow \chi  \bar \chi$. To estimate the overall meson yield, we assume 2.4 pions are produced per POT~\cite{Aguilar-Arevalo:2018wea}, while the production of $\eta$ ($\eta'$) is smaller than pion production by a factor of $30$ $(300)$~\cite{deNiverville:2016rqh}. We have simulated the production of  $\eta(')$ mesons at the Booster with PYTHIA 8~\cite{Sjostrand:2014zea}. These events are passed to a dedicated simulation which first decays the mesons to dark matter particles and then estimates the probability of passing through the detector and scattering. The MiniBooNE detector, consisting of 800 tonnnes of mineral oil, was located 491 meters downstream of the beam dump. The differential cross section for dark matter - nucleon elastic scattering in our scenario is given by
\begin{equation}
\frac{d \sigma_{\chi N} }{d Q^2} = \frac{g_\chi^2 \, y_{SNN}^2 }{64 \pi m_N^2} \frac{(4 m_\chi^2+Q^2)(4 m_N^2+Q^2)} { (E^2-m_\chi^2) (m_S^2 + Q^2)^2 }F^2(Q^2), 
\label{eq:form-factor}
\end{equation}
where $Q^2 = 2 m_N T$, with $m_N$ the nucleon mass and $T$ the nucleon recoil kinetic energy. 
The effective scalar-nucleon couplings in Eq.~(\ref{eq:form-factor}) are given by (see Eq.~(\ref{eq:ySNN}) in the Appendix)
\begin{eqnarray}
y_{Spp} & = &  g_u \, \langle p | \bar u u |p \rangle  = g_u \frac{f_{Tu}^p m_p}{m_u}  \approx 6.0 \, g_u, \nonumber \\
y_{Snn} & = &  g_u \, \langle n | \bar u u |n \rangle  = g_u \frac{f_{Tu}^n m_n}{m_u}  \approx 5.1 \, g_u,
\label{eq:S-nucleon-couplings}
\end{eqnarray}
where in the last equality we have taken  $f_{Tu}^p = 0.014$,  $f_{Tu}^n = 0.012$~\cite{Durr:2015dna}.
We have also assumed a dipole form factor $F(Q^2) = (1+Q^2/M^2)^{-2}$, with $M^2 = 0.55~{\rm GeV}^2$~\cite{Schweitzer:2003sb}. 
The DM search was performed in the  nucleon recoil kinetic energy window of $35\,{\rm MeV}\,<T<\,600\,{\rm MeV}$. 
We model the detector effects with a simple step function, applying a 35$\%$ detection efficiency for events with $T  > 100\,{\rm  MeV}$ and cutting events with smaller recoil energies. After all selections,  $1465\pm38$ events were observed, while $1548\pm198$ background events were expected. With the data and background predictions in agreement, a 90$\%$ CL limit is derived by demanding the number of dark matter scattering events is less than $1.64 \sqrt{1548+(198)^2} \simeq 331$. We have checked that this procedure reproduces well the limits derived in the vector portal DM model~\cite{Aguilar-Arevalo:2017mqx}. The parameter space in our model limited by this search is shown in the right panel of Fig.~\ref{fig:suulimits}.   
We have also estimated the sensitivity of a possible future dedicated beam dump experiment using the Fermilab booster and the SBND detector~\cite{vandewater}. The projection, also shown in Fig.~\ref{fig:suulimits}, is made by rescaling the MiniBooNE-DM limit accounting for a factor of $\sim 20$ reduction in the expected neutrino flux with a dedicated dump and the different geometric acceptance and detector mass of SBND.

\subsection{BBN}

If $S$ is in thermal contact with the SM at the time of BBN, it contributes to the number of effective relativistic degrees of freedom at this time, which is tightly constrained by primordial element abundances. 
Furthermore, the decays of $S$ to photons for $m_S < 2 m_\pi$ affects the baryon-to-photon ratio, which has the net effect of increasing the deuterium abundance $\text{D/H} = (2.53 \pm 0.04)\times 10^{-5}$ \cite{Cooke:2013cba}. Recasting the bounds of~\cite{Millea:2015qra}, we find the conditions
\be
\begin{split}
m_S &> 20\text{ MeV},\\
g_u &> (2\times 10^{-8})\Bigl( \frac{m_S}{\text{GeV}}\Bigr)^{-3/2},
\end{split}
\ee
where the first bound is from the case where the scalar generally stays in thermal equilibrium until kinetic decoupling, while the second bound describes the limit from scalars which decay before BBN but still affect the baryon-to-photon ratio through late decays to photons. 

The results of Ref.~\cite{Millea:2015qra} do not apply if $S$ has never reached thermal equilibrium throughout the entire history of the universe, which may occur when $g_u$ becomes very small. While a detailed analysis of the thermalization condition is rather involved (see e.g.\ Ref.~\cite{Salvio:2013iaa}) and beyond the scope of this article, we can obtain a rough estimate from a dimensional argument: Before the QCD phase transition, the main scalar production processes are $u\bar{u} \to S$, $u\bar{u} \to Sg$ and $ug \to uS$.
For small $m_S$ the $S$ production rate is proportional to $T$, $\Gamma_S \sim g_u^2T$. The scalars would reach thermal equilibrium when $\Gamma_S$ is greater than the Hubble rate $H \sim \sqrt{g^*}T^2/M_{\rm Pl}$. Since the number of degrees of freedom, $g^*$, drops sharply at the QCD phase transition, we only consider $T>1$~GeV. Then the thermalization condition leads to the requirement $g_u \gtrsim 10^{-9}$.

For $m_S > 2 m_\pi$ and the range of couplings considered here ($g_u \gtrsim 10^{-10}$),
there are no relevant bounds from BBN since $S$ decays well before one second.

For an invisibly decaying scalar, there is in principle a bound from $N_{\mathrm{eff}}$ if there are appreciable densities of $S$ and $\chi$ during BBN. 
However, it is difficult to put a meaningful model-independent constraint from BBN because if $S$ is ever in thermal equilibrium with the SM and decays to $\chi\bar{\chi}$, generally dark matter is overproduced in the early universe assuming standard thermal cosmology. 
Thus, one would have to extend the model to rectify this issue. For instance, one could add couplings of $S$ to electrons and/or neutrinos or introduce an extra light degree of freedom, which would provide a means to deplete the $S$ abundance~\cite{Knapen:2017xzo}. 
The resulting influence on BBN would then depend on the new interactions and states in the model, and so we do not show an explicit limit here.

\subsection{Supernovae}

Emission of very weakly coupled scalars could have led to increased energy loss from the core of SN 1987A, in conflict with the observed light curve.
The leading contribution for the up-philic scalar $S$ stems from the 
bremsstrahlung process $N N \to N N + S$. Following Ref.~\cite{Hanhart:2000er}, we treat the nucleons in a non-relativistic approximation, leading to a factorized differential cross-section
\begin{align}
d\sigma[NN \to NN+S] &\approx d\sigma[NN \to NN] \; \frac{d^3k_S}{(2\pi)^3
2E_S}\beta_{\rm f} \; y_{SNN}^2 \notag \\
&\qquad \times \biggl [
 \frac{2E_S^3-m_S^2E_S -2m_S [({\rm\bf k}_S\cdot{\pmb\beta}_{\rm i})^2 -
 ({\rm\bf k}_S\cdot{\pmb\beta}_{\rm f})^2]}{m_NE_S^3}
\biggr ]^2
\end{align}
where $k_S$ and $E_S$ are the 3-momentum and energy of the final-state $S$,
respectively, and $y_{SNN}$ is the effective nucleon-$S$ Yukawa coupling.
Furthermore, $\beta_{\rm i,f}$ are the non-relativistic velocities of an
initial-state and a final-state nucleon, respectively. They are connected via
the relationship $E_S = m_N(\beta_{\rm i}^2-\beta_{\rm f}^2)$.

The thermally averaged energy loss rate is given by \cite{Raffelt:1990yz}
\be
Q_S(T) = \int_{\sqrt{m_S/m_N}}^\infty d\beta_{\rm i}\,\beta_{\rm i} \; f_N(T,\beta_{\rm i})
\int d\sigma[NN \to NN+S] \; E_S \, n_N^2,
\ee
where $n_N \approx 1.8 \times 10^{38}\,\text{cm}^{-3}$ is the nucleon number density in the SN core, and $f_N$ is the Maxwellian nucleon distribution function. By imposing the observational bound \cite{Raffelt:1990yz}
\be
Q_S \lesssim 3 \times 10^{33} {\rm\ erg\ cm}^{-3} {\rm\ s}^{-1},
\ee
and using $T \approx 30\,$MeV and $\sigma{[NN\to NN]} \approx 2.5\times 10^{-26}\,\text{cm}^2$ \cite{Hanhart:2000er}, one obtains a bound on $y_{SNN}$. Our results are consistent, within a factor of two, with the analysis of Ref.~\cite{Ishizuka:1989ts}. The bound on $y_{SNN}$ can be translated to a bound on $g_u$ by using the relations \eqref{eq:S-nucleon-couplings}. Since $f^p_{Tu}$ and $f^n_{Tu}$ differ only by ${\cal O}(20\%)$, we assume for simplicity that the SN core plasma contains only neutrons, i.e.\ we identify $y_{SNN}\approx y_{Snn}$. The resulting limit is shown by the lower edge of the region labeled ``SN 1987A'' in Fig.~\ref{fig:suulimits}.

For sufficiently large $g_u$ the scalars emitted through bremsstrahlung are trapped within the supernova core and cannot escape. The trapping can occur through absorption, $NN+S \to NN$, or decay of $S$. The former can be evaluated by considering the effective mean opacity \cite{Raffelt:1988rx,Raffelt:1996wa}
\begin{align}
\kappa_S = \frac{\frac{8}{15}\pi^4T^3}%
{m_N\int_{m_S}^\infty dE_S \; E_S^3\,(1-m_S^2/E_S^2) \, \sigma_{\rm abs}^{-1} \;
\frac{\partial}{\partial T} (e^{E_S/T}-1)^{-1}}
\label{eq:kappas}
\end{align}
where $\sigma_{\rm abs}$ is the thermally averaged absorption cross-section, which can be computed by again using the non-relativistic approximation.

The decay contribution is only relevant if the decay length is smaller than the size of the SN core. In our case this requires $m_S > 2m_\pi$, which is too massive for efficient scalar production in the SN. For completeness, we nevertheless include the decay contribution to the opacity in our numerical evaluation.

A significant fraction of the $S$ scalars will be trapped inside the SN core
if $\kappa_S$ is greater than the neutrino opacity $\kappa_\nu$ \cite{Dolan:2017osp}, where
$\kappa_\nu \approx 8\times 10^{-17}\, \text{cm}^2/\text{g}$ \cite{Masso:1995tw}.
This leads to the upper edge of the exclusion region shown in
Fig.~\ref{fig:suulimits}~(top left).

The trapping bound changes for invisibly decaying scalars. Assuming that the decay $S \to \chi\bar{\chi}$ is prompt (i.e.\ $g_\chi$ is sufficiently large), this constraint is governed by the $\chi+N$ scattering rate rather than the absorption rate of $S$. We will conservatively assume that a single scattering event is sufficient to trap a dark matter particle inside the SN core. Then the opacity $\kappa_\chi$ can be computed by simply replacing $\sigma_{\rm abs}$ in Eq.~\eqref{eq:kappas} with the thermal average of $\sigma_{\chi N}$ from Eq.~\eqref{eq:form-factor}. The resulting bound is given by the upper edge of the ``SN 1987A'' region shown in
Fig.~\ref{fig:suulimits}~(top right).

\subsection{Direct detection}

Conventional direct detection experiments searching for dark matter induced nuclear recoils can probe dark matter at or above the GeV scale. The effective spin-independent DM-nucleon scattering cross section is
\begin{equation}
\sigma^{\rm SI}_{\chi N} = \frac{\mu_{\chi N}^2}{\pi}\frac{(Z f_p +(A-Z)f_n)^2}{A^2},
\end{equation}
where $Z$ ($A$) are the atomic (mass) number of the nuclear target, $\mu_{\chi N}$ is the DM-nucleon reduced mass, and $f_p, f_n$ are the effective DM - nucleon couplings, 
$f_N  = g_\chi y_{SNN} / m_S^2$. The scalar-nucleon effective couplings were discussed above in Eq.~(\ref{eq:S-nucleon-couplings}) (see also Eq.~(\ref{eq:ySNN}) in the Appendix).

Given an assumption on the mass ratio $m_\chi / m_S$ and the dark coupling $g_\chi$, direct detection bounds may be shown on the $m_S$--$g_u$ plane. In the cases where DM annihilates to SM particles, we assume $g_\chi = 1$ and choose $m_\chi = (3/4) m_S$ ($m_\chi = (1/3) m_S$) as a benchmark with a visibly (invisibly) decaying scalar. Then, direct detection constraints may be shown for any choices of $m_S$ and $g_u$, as indicated on Fig.~\ref{fig:suulimits}. There is a ``thermal DM'' target line where annihilation $\chi \bar{\chi} \to \mathrm{SM}~\mathrm{SM}$ produces the observed amount of DM. Away from this region, a non-thermal mechanism is necessary to set the DM relic abundance, or if $\chi$ makes up a subdominant component of the DM (above the thermal target band), the direct detection limits which we have shown would have to be rescaled. When the relic abundance is set through secluded annihilation, choosing the scalar and DM mass immediately implies a specific DM coupling $g_\chi$, independently of $g_u$. Therefore, we also show the direct detection limits for a secluded annihilation benchmark with $m_\chi = 3 m_S$. We emphasize that unlike in the visible annihilation case, the observed relic density can be obtained through secluded annihilation at every point in the $m_S$--$g_u$ plane.

The current direct detection limits are a combination of the $\nu$-cleus~\cite{Angloher:2017sxg}, CRESST~\cite{Angloher:2015ewa,Petricca:2017zdp}, CDMSlite~\cite{Agnese:2015nto}, PICO~\cite{Amole:2017dex}, and XENON1T~\cite{Aprile:2017iyp} experiments. We also show the expected performance~\cite{Battaglieri:2017aum} of one future direct detection search aimed at observing low energy recoils, the NEWS-G experiment \cite{Arnaud:2017bjh} which uses light gaseous targets. Superfluid helium detector concepts in the research and development phase may probe even lower DM masses~\cite{Guo:2013dt,Carter:2016wid,Maris:2017xvi}.

\subsection{Colliders}

Finally, an invisibly decaying $S$ can lead to jet plus missing energy events at the LHC through $q \bar{q} \to S g, S \to \bar{\chi} \chi$. We perform a simple parton-level recasting of the current ATLAS monojet search~\cite{Aaboud:2017phn} using MadGraph 5~\cite{Alwall:2014hca} and a modified version of a simplified model for DM coupling to a scalar mediator~\cite{Mattelaer:2015haa}. The resulting limit is $g_u \lesssim 0.1$ and is independent of the scalar mass within the range we consider, as we are only considering $m_S$ much below the typical energies in the events for which ATLAS searches, on the order of hundreds of GeV of $p_T$.

\subsection{Summary}

\begin{figure}
\centering
\includegraphics[width=0.49\textwidth]{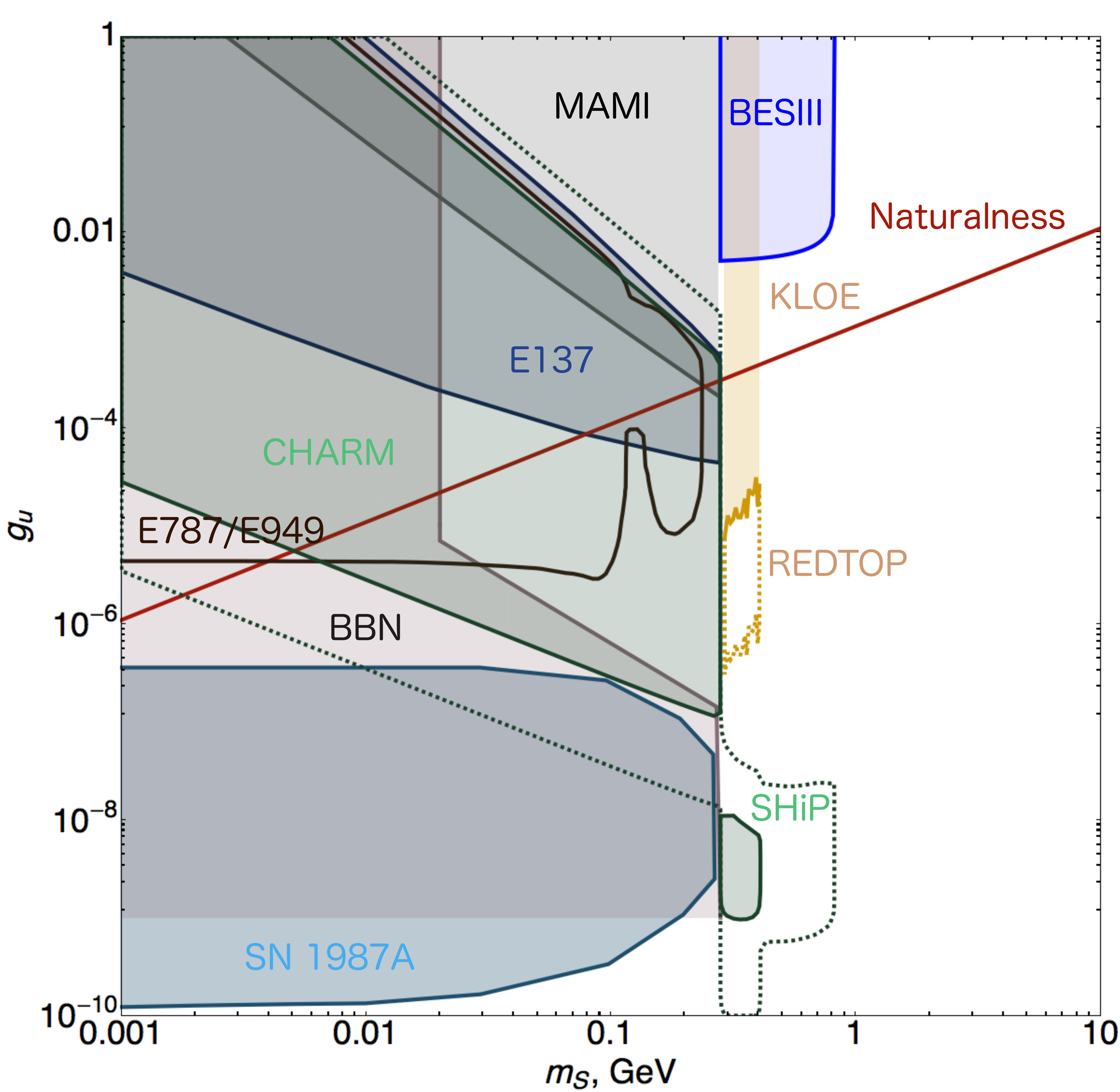}
\includegraphics[width=0.49\textwidth]{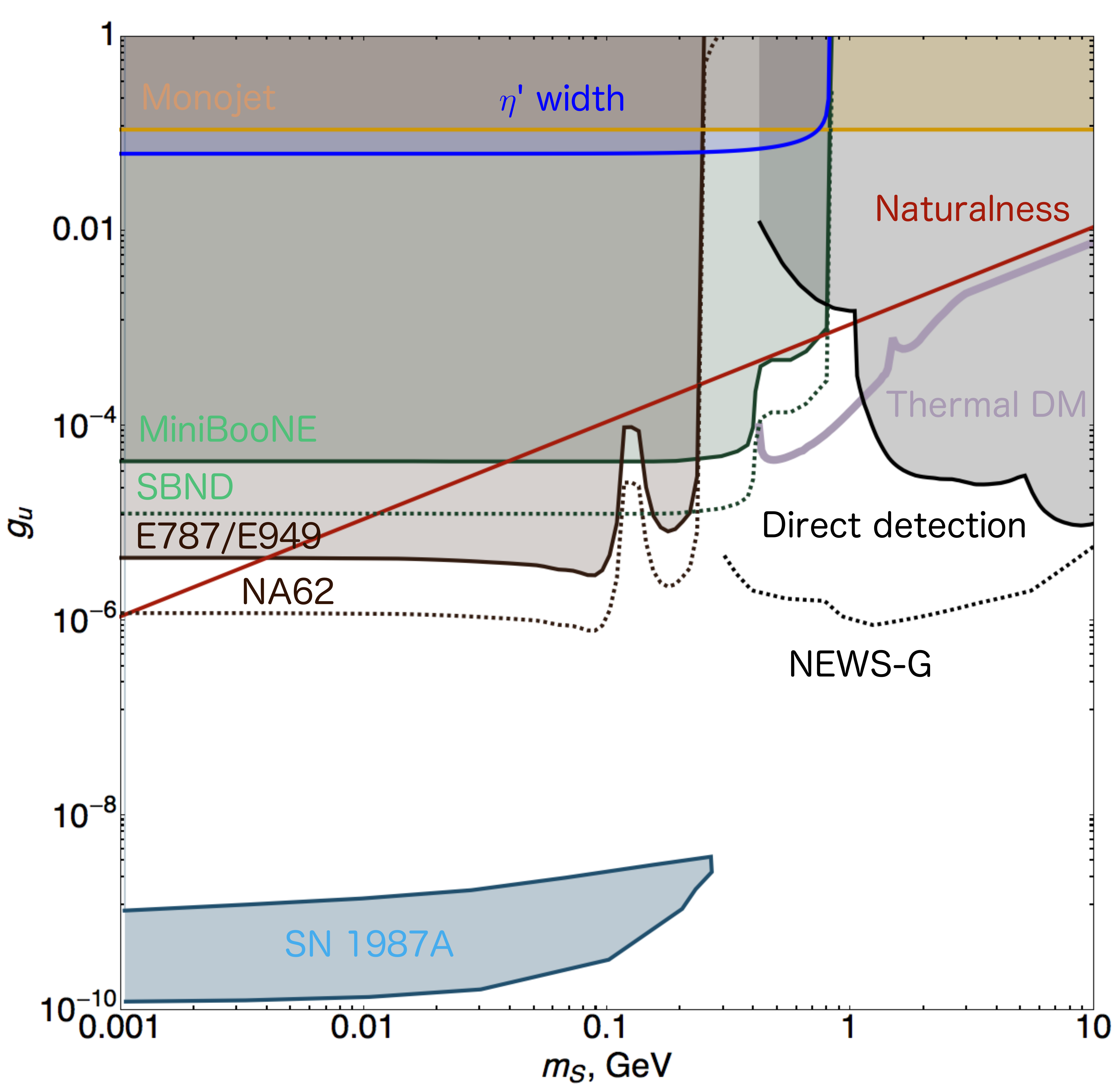}
\includegraphics[width=0.49\textwidth]{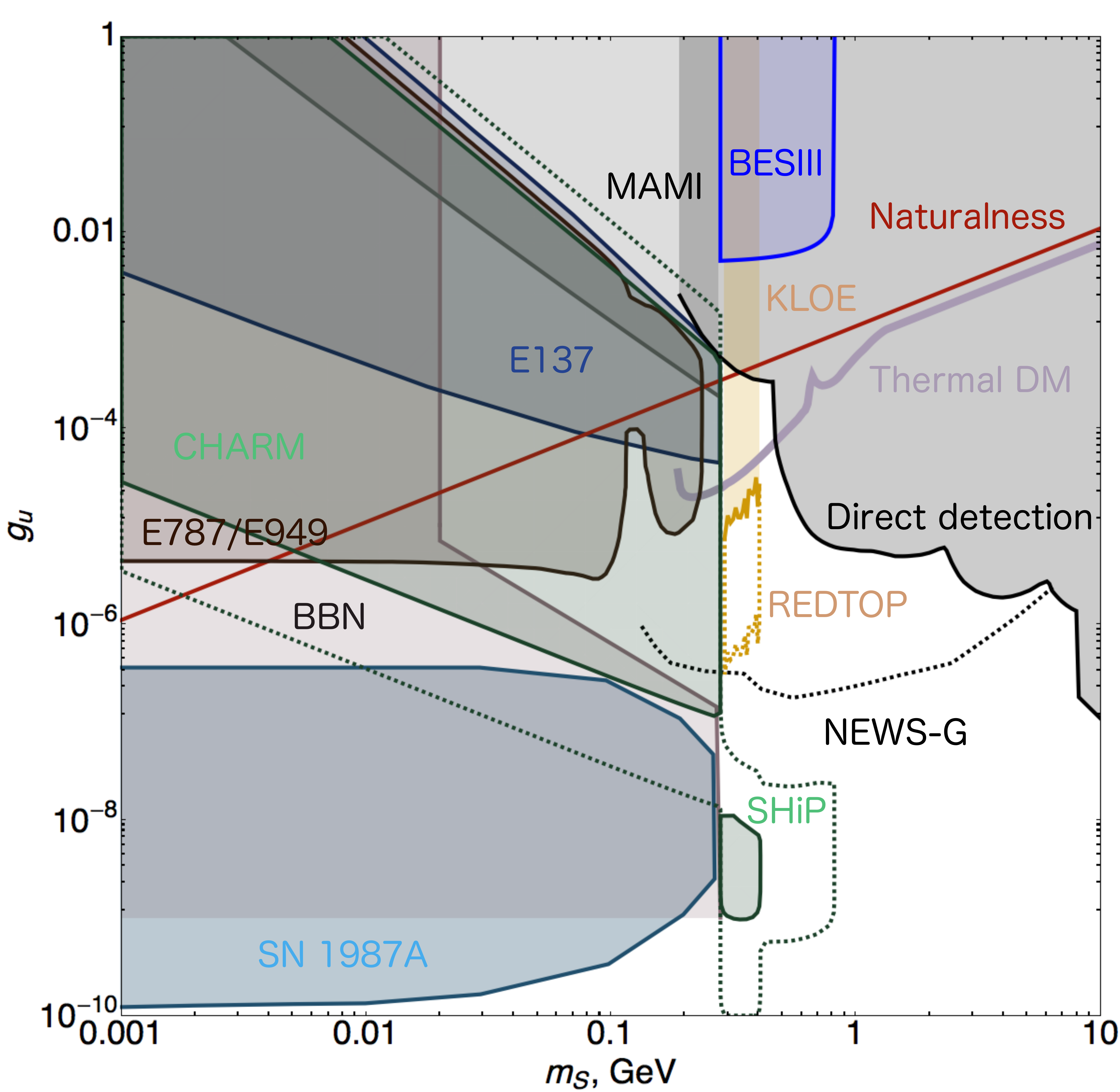}
\includegraphics[width=0.49\textwidth]{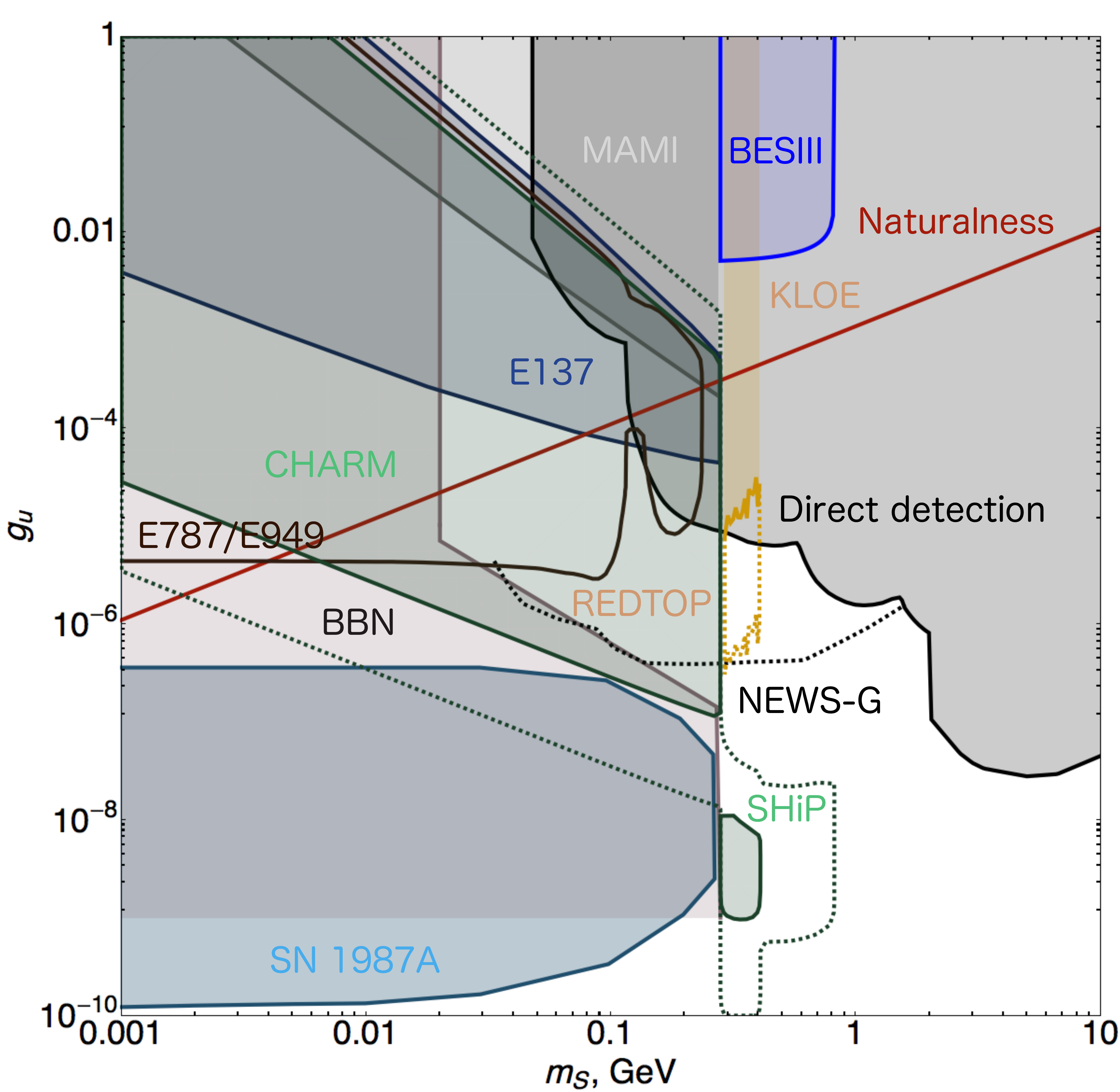}
\caption{\label{fig:suulimits}
Top left: Limits on an up-philic scalar mediator. Top right: Limits on an invisibly decaying up-philic scalar. The thermal dark matter and direct detection lines are drawn assuming $m_\chi = (1/3) m_S$ and $g_\chi = 1$. Bottom left: Same as top right panel but for a visibly decaying scalar, with all dark matter lines assuming $m_\chi = (3/4) m_S$. Bottom right: Limits on the up-philic scalar with $m_\chi = 3 m_S$. The DM coupling $g_\chi$ is chosen at each point such that secluded annihilation sets the correct thermal relic density.
Note that some of the bounds are at 90\% CL (direct detection, E787/E949, MiniBooNE) or 95\% CL (BBN, CHARM, SHiP, MAMI, KLOE), while the SN 1987A exclusion region has no defined confidence level.
}
\end{figure}
We summarize the limits in this section in Fig.~\ref{fig:suulimits}. We show both the limits on the scalar mediator $S$ itself, as well as those on the scenario where the scalar mediates interactions between the SM and DM under various assumptions on the DM mass and coupling. 

First, the top left panel of Fig.~\ref{fig:suulimits} contains bounds on $S$, assuming no additional scalar decay modes other than those to SM particles. Below the pion threshold, the scenario is highly constrained by a combination of precision meson decay measurements, fixed-target experiments, and BBN. At very low couplings such that $S$ is never in thermal contact with the SM, there may be additional viable parameter space, and it would be interesting to examine this region further.

The top right and bottom left panels of Fig.~\ref{fig:suulimits} contain our results for DM annihilating to SM particles through the scalar mediator, where $S$ decays to DM or SM particles, respectively. For the invisibly decaying $S$ in the top right panel, we see that for $m_S$ between $\sim 1$ and hundreds of MeV, rare meson decays prove to be constraining, but there is a gap between their reach and the region probed by SN 1987A. This represents an interesting area that we hope will be probed by future experiments. At higher masses, the situation is less clear, as the position of the dark matter target and constraint lines are highly dependent on the ratio $m_\chi / m_S$, which is arbitrary except that $S \to \bar{\chi} \chi$ must be kinematically accessible for $S$ to decay invisibly. The most robust bound is perhaps the LHC monojet search, which for these scalar masses does not probe the region of parameter space that is plausible from the point of view of technical naturalness. The GeV range of this model thus also contains a useful potential target. Our limits stand in contrast to those on a Higgs-like scalar, where since the couplings of the Higgs and scalar are related by a single mixing angle, invisible Higgs decays universally provide a strong constraint as long as $m_\chi < m_h / 2$.

Conversely, for the visible annihilation case in the bottom left panel where $S$ decays visibly, the DM-related limits would not move significantly as $m_\chi / m_S$ is varied within its natural range of 1/2 to 1. The region below the pion threshold is well limited, and there is only a small window above the pion threshold for which the thermal DM scenario is viable, owing to a combination of direct detection and the KLOE measurement of $\eta \to 3 \pi$. Here, NEWS-G shows sensitivity to the remaining sliver of the thermal target parameter space.

The bottom right panel of Fig.\ref{fig:suulimits}, finally, shows the limits on a dark sector containing DM annihilating directly to a scalar up-philic mediator. The positions of the DM lines are affected by our assumption on $m_\chi / m_S$, which is taken to be 3 in the plot, but we anticipate that the overall qualitative picture would not change if this ratio were to be varied. Generally, there is open space for $m_S$ above the pion threshold, in the region below the direct detection limits. Future direct detection experiments such as NEWS-G, and to some extent beam-dump experiments such as SHiP, will be sensitive to this area.

We have also looked at constraints stemming from other sources, such as neutron scattering and electroweak precision data, but found them to be not competitive with the bounds shown here.

\section{Conclusions}
\label{sec:concl}
% !TEX root = up-dm.tex

Light dark sectors are a particularly interesting realm of contemporary BSM phenomenology, with promising precision, beam dump, and direct detection experiments on the horizon. 
The standard renormalizable portals involving a dark Higgs, dark photon or sterile neutrino have been well-examined in the literature, but much work remains to be done in identifying and investigating viable models with even broader phenomenology. In this work we have conducted a study of one such model
%, which can be considered a generalization of the Higgs portal scenario, 
in which a scalar mediator preferentially couples to a light SM quark, leading to a hadrophilic dark sector with a distinctive phenomenology. 
We have identified the viable regions of parameter space where the dark matter abundance is set through thermal freeze-out and have performed an extensive analysis of the current experimental constraints and future probes of this scenario. 

The scalar mediator we have considered is flavor-aligned but technically natural, and it represents an interesting complementary benchmark to a Higgs-like scalar. In contrast to the latter, many of the strongest bounds from rare heavy meson decays are avoided, through the absence of penguin diagrams involving top quarks with mediator emission. Conversely, relatively large up quark couplings are allowed in our model, in contrast to the usual Higgs-like scalar mediator case where the largest reasonable $S\bar{u}u$ coupling is of order $m_u / m_t$. This allows for significant production of scalars in \emph{light}-flavor meson decays. In general, $\eta$ decays at proton fixed-target experiments provide a significant source of scalars with mass below a few hundred MeV. We have examined the resulting constraints and prospects from current and future proton beam dumps, as well as precision $\eta$ measurements. At somewhat higher masses the $\eta'$ provides some sensitivity as well. Whether the scalar decays visibly or invisibly, relevant limits can be obtained from meson decays. Future $K$ and $\eta$ measurements at experiments such as NA62 and REDTOP, respectively, can improve these limits considerably.

In the case that the new scalar decays to SM particles, it is relatively straightforward to derive astrophysical and cosmological constraints at low $m_S$ from BBN and supernova considerations. If the scalar has a significant invisible branching fraction into dark matter, there is an inherent link between this decay channel and the dark matter relic abundance. However, this link is highly model-dependent since the relic density may be affected by other states in the dark sector. Thus we have chosen to analyze constraints from BBN and supernovae independent of the mechanism that may set the dark matter abundance. It would be interesting to study the relation between these in more detail in the future. 
Meanwhile, at GeV-scale mediator masses direct detection starts to play an important role. In this regard, it is interesting that future low threshold experiments such as NEWS-G offer the possibility to cover significantly wider swaths of parameter space, notably including the remaining allowed thermal dark matter target region for a visibly decaying scalar.

While we have focused on a scalar mediator that dominantly couples to the up-quark, the alignment hypothesis from which we have started could be applied equally to any of the quarks, or indeed to a single up- and down-type quark simultaneously. For instance, in the case of a down-philic scalar, much of the phenomenology would be identical to the case which we have considered here, due to strong isospin. One notable difference would be that while an up-philic scalar couples to charged kaons, a down-philic scalar couples only to neutral kaons, which could have an impact on some of the precision meson limits which we have studied. It would be interesting to pursue this possibility, and other flavor-specific cases, in more detail in the future. 

It is known that the theory space for new light scalar mediators is broader than the simple Higgs portal case. Here, we have performed a case study of such a model, demonstrating the differences with respect to a standard Higgs-like scalar. As the exploration of light dark sectors comes of age, we should continue to investigate a broad range of viable and well-motivated theories that can provide interesting and diverse phenomenological signatures.

\acknowledgments

We thank Tyler Thornton, Sean Tulin, and Richard Van De Water for useful discussions. 
The work of BB and AI is supported in part by the U.S.~Department of Energy under grant No.~DE-SC0007914. The work of AF is supported in part by the National Science Foundation under grants No.~PHY-1519175 and No.~PHY-1820760. DM is supported by the National Research Council of Canada. This work was performed in part at Aspen Center for Physics, which is supported by National Science Foundation grant PHY-1607611. AI would like to thank the Galileo Galilei Institute for Theoretical Physics for the hospitality and the INFN for partial support during the completion of this work.

\appendix
\numberwithin{equation}{section}

\section{Hadronic couplings for a general flavor-diagonal scalar}
\label{app:ChiralL}
% !TEX root = up-DM.tex

%\documentclass[12pt]{article}
%\topmargin -.8cm
%%\textwidth=14.5cm
%%\textheight=20.5cm
%\oddsidemargin  10.5pt
%\evensidemargin  10.5pt
%\textheight  612pt
%\textwidth  432pt
%%\headheight  12pt
%%\headsep  20pt

%\usepackage{amsmath,amsfonts,amssymb,graphicx}
%\usepackage{epsfig}
%\usepackage{amsmath}
%\usepackage{amsfonts}
%\usepackage{tabularx}
%\usepackage{bigstrut}
%\usepackage{multirow}
%\usepackage{slashed}
%\usepackage{array}
%\usepackage{bbold}
%\usepackage[table]{xcolor}
%\usepackage{cancel}

%\def\BB#1{{\bf  \textcolor{blue}{BB: {#1}}}}

%\linespread{1.2}

%\begin{document}

%%%%%%%%%%%%%%%%%%%%%%%%
%%%%%%%%%%%%%%%%%%%%%%%%
%\section{General flavor-diagonal scalar}

We consider a new singlet scalar particle $S$. We assume the scalar interacts with quarks and gluons through the Lagrangian
\begin{equation}
{\cal L} = - \sum_{\psi} \kappa_\psi \frac{m_\psi}{v} S\, \bar \psi \, \psi + \kappa_G \, \frac{\alpha_s}{12 \pi} S \, G_{\mu\nu}^a G^{\mu\nu a},
\label{eq:L1}
\end{equation}
where the sum runs over all SM quarks $\psi = u,d,s,c,b,t$. This Lagrangian is defined at scales of order 100 GeV, and we consider general coupling coefficients $\kappa_\psi, \kappa_G$ induced by new physics above this scale. We note that the SM Higgs couplings correspond to $\kappa_u = \kappa_d = \kappa_s = \kappa_c = \kappa_b = \kappa_t = 1$, $\kappa_G = 0$. For the case of an up-philic scalar studied in this work, the Lagrangian of Eq.~(\ref{eq:bsmlag}) implies $\kappa_u = g_u v / m_u$ and $\kappa_d = \kappa_s = \kappa_c = \kappa_b = \kappa_t = \kappa_G = 0$.

At low scales of order 1 GeV we integrate out the heavy quarks $c,b,t$ to obtain the effective Lagrangian
\begin{equation}
{\cal L} =  \frac{S}{v} \left[ ( \kappa_c + \kappa_b+ \kappa_t + \kappa_G )  \, \frac{\alpha_s}{12 \pi} G_{\mu\nu}^a G^{\mu\nu a} - \kappa_u \, m_u  \bar u u - \kappa_d \, m_d  \bar d d  - \kappa_s \, m_s \bar s s  \right]. 
\label{eq:L2}
\end{equation}
We write these interactions in terms of the trace of the energy-momentum tensor, which is given by
\begin{equation}
\Theta^\mu_\mu = -\frac{9 \alpha_s}{8 \pi}  G_{\mu\nu}^a G^{\mu\nu a} +  m_u \bar u u +  m_d  \bar d d  +  m_s \bar s s.
\end{equation}
The Lagrangian (\ref{eq:L2}) becomes
\begin{equation}
{\cal L} = - \frac{S}{v} \left[ \frac{2}{9} K_\Theta \, \Theta^\mu_\mu + \frac{7}{9} \left(K_u \, m_u  \bar u u + K_d \, m_d  \bar d d  + K_s \, m_s \bar s s  \right) \right]. 
\label{eq:L3}
\end{equation}
where we have defined
\begin{eqnarray}
K_\Theta =  \frac{1}{3}( \kappa_c + \kappa_b+ \kappa_t + \kappa_G ), ~~~~~~&~~~& K_u = \frac{9}{7}\left[\kappa_u -\frac{2}{27}( \kappa_c + \kappa_b+ \kappa_t + \kappa_G )\right], \nonumber \\
K_d = \frac{9}{7}\left[\kappa_d -\frac{2}{27}( \kappa_c + \kappa_b+ \kappa_t + \kappa_G )\right], & ~~~& K_s = \frac{9}{7}\left[\kappa_s -\frac{2}{27}( \kappa_c + \kappa_b+ \kappa_t + \kappa_G )\right].~~~~~~~
\end{eqnarray}
In the limit of the SM Higgs couplings, we have $K_\Theta = K_u = K_d = K_s = 1$. For the up-philic scalar considered in the main text, we have $K_u = (9/7) g_u v /m_u$ and $K_\Theta = K_d = K_s = 0$.

\subsection{Matching to the chiral Lagrangian}

We describe the Goldstone bosons (including the $\eta'$) with the $\Sigma$ field, 
\begin{equation}
\Sigma(x) = e^{2 i \pi(x)/f}.
\label{eq:xi-Sigma}
\end{equation}
The pion matrix is parameterized as $ \pi = \pi^a t^a + \eta_0 t^0$, where $t^a$ are the usual $SU(3)$ generators and the $U(1)$ generator is $t^0 = \tfrac{1}{\sqrt{6}}  \mathbb{1}$. The explicit form of the pion matrix is 
\begin{equation}
 \pi =
\renewcommand\arraystretch{1.4}
\frac{1}{\sqrt{2}}   \begin{pmatrix}
 \tfrac{1}{\sqrt{2}} \pi^0 +  \frac{1}{\sqrt{6}}\eta_8  +  \frac{1}{\sqrt{3}}\eta_0   & \pi^+ & K^+  \\
\pi^- &   -\tfrac{1}{\sqrt{2}} \pi^0 +  \frac{1}{\sqrt{6}}\eta_8 +  \frac{1}{\sqrt{3}}\eta_0    & K^0  \\
 K^- &  \overline K^0 & - \tfrac{2}{\sqrt{6}} \eta_8 +   \frac{1}{\sqrt{3}}\eta_0  
\end{pmatrix}.
    \label{eq:pion-parameterization}
\end{equation}
The leading terms in the Lagrangian are 
\begin{equation}
{\cal L}  \supset  \frac{f^2}{4}   {\rm tr} [ \,D_\mu \Sigma \, D^\mu \, \Sigma^\dag \, ] +  B \frac{f^2}{2}  \left\{ {\rm tr}\left[ \Sigma^\dag  m_q \right]  + {\rm h.c.} \right\}
- a \frac{f^2}{4 N_c} \left( -i \ln \det \Sigma  \right)^2.
\label{eq:L4}
\end{equation}
where $m_q = {\rm diag}(m_u, m_d, m_s)$ is the quark mass matrix. The last term in Eq.~(\ref{eq:L4}) captures the effect of the anomaly and leads to a mass term for the $\eta_0$~\cite{Witten:1980sp}, given by $m_0^2 = 3 a/N_c$.

We now wish to match the quark and gluon interactions to the Goldstone interactions, as a prelude to matching the interactions of a new scalar with quarks and gluons to scalar-meson couplings. To do this, we use the relations
\begin{equation}
\Theta^\mu_\mu =  \frac{f^2}{2}   {\rm tr} [ \,D_\mu \Sigma \, D^\mu \, \Sigma^\dag \, ]  - 4 {\cal L}, ~~~~~~~~~ m_q \, \bar q q = - m_q \, \frac{\partial {\cal L}}{\partial m_q}.
\label{eq:matching}
\end{equation}
%At the end of the computation, we also take the isospin limit by making the replacements
in conjunction with Eq.~(\ref{eq:L4}) to match between bilinears involving the elementary quark fields/gluon field strength and those involving the mesons. The chiral Lagrangian also contains meson masses, motivating the (isospin limit) replacements
\begin{equation}
B m_u \rightarrow \frac{1}{2}m_\pi^2 , ~~~~~~~ B m_d \rightarrow \frac{1}{2}m_\pi^2 , ~~~~~~~ B m_s \rightarrow m_K^2 - \frac{1}{2} m_\pi^2. ~~~~
\end{equation}
%To quadratic order in the pion fields, we find
%\begin{eqnarray}
%\Theta^\mu_\mu & = & -2 |\partial_\mu \pi^+|^2 - (\partial_\mu \pi^0)^2 - 2 |\partial_\mu K^+|^2 -2 |\partial_\mu K^0|^2 - (\partial_\mu \eta_8)^2 - (\partial_\mu \eta_0)^2 \nonumber \\
%& &+ 4 m_\pi^2 |\pi^+|^2 + 2 m_\pi^2 (\pi^0)^2 + 4 m_K^2 |K^+|^2  + 4 m_K^2 |K^0|^2  +\frac{2}{3}(4m_K^2-m_\pi^2) \eta_8^2  \nonumber \\
%&& + \left[ \frac{2}{3}(2m_K^2+m_\pi^2)  + 2 m_0^2 \right] \eta_0^2  + \frac{8 \sqrt{2}}{3} (m_\pi^2 - m_K^2)\eta_8 \eta_0, \nonumber \\
%m_u \bar u u + m_d \bar d d & = & m_\pi^2 |\pi^+|^2 + \frac{1}{2} m_\pi^2 (\pi^0)^2 + \frac{1}{2}m_\pi^2 |K^+|^2 + \frac{1}{2}m_\pi^2 |K^0|^2 \nonumber \\
%&& + \frac{1}{6} m_\pi^2 (\eta_8)^2
%+ \frac{1}{3} m_\pi^2 (\eta_0)^2 + \frac{\sqrt{2}}{3} m_\pi^2 \eta_8 \eta_0 \nonumber \\
%m_u \bar u u - m_d \bar d d & = &  \frac{1}{2}m_\pi^2 |K^+|^2 - \frac{1}{2}m_\pi^2 |K^0|^2 + \frac{1}{\sqrt{3}} m_\pi^2 \pi^0 \eta_8 + \sqrt{\frac{2}{3}} m_\pi^2 \pi^0 \eta_0. \nonumber \\
%m_s \bar s s  & = &  \left(m_K^2-\frac{1}{2}m_\pi^2  \right) |K^+|^2 +\left(m_K^2-\frac{1}{2}m_\pi^2  \right) |K^0|^2 +\frac{2}{3}\left(m_K^2-\frac{1}{2}m_\pi^2  \right) (\eta_8)^2  \nonumber \\ 
%&& +\frac{1}{3} \left(m_K^2-\frac{1}{2}m_\pi^2  \right) (\eta_0)^2 -\frac{2\sqrt{2}}{3} \left(m_K^2-\frac{1}{2}m_\pi^2  \right)\eta_8 \eta_0.
%\label{eq:matching2}
%\end{eqnarray}
%Inserting these expressions into the Lagrangian (\ref{eq:L3}) gives the cubic couplings of the scalar $S$ to two Goldstone bosons.

We describe the result of this matching by writing interactions of the scalar to the $\Sigma$ field in the chiral Lagrangian. The leading terms in the Lagrangian are 
\begin{eqnarray}
{\cal L}  &\supset &  \frac{f^2}{4} \left(1+ c_1 \frac{S}{v} \right)   {\rm tr} [ \,D_\mu \Sigma \, D^\mu \, \Sigma^\dag \, ] + 
 B \frac{f^2}{2}\left\{ {\rm tr} \left[\Sigma^\dag (m_q + \frac{S}{v} c_q) \right]  + {\rm h.c.}  \right\} \nonumber \\
&&-  \left(1+ c_2 \frac{S}{v} \right) a \frac{f^2}{4 N_c} \left( -i \ln \det \Sigma  \right)^2.
\label{eq:L5}
\end{eqnarray}
where we have introduced the coupling matrix $c_q = {\rm diag}(c_u m_u, c_d m_d, c_s m_s)$. The matching approach derived in Eqs.~\ref{eq:matching} and \ref{eq:L3} yields
\begin{equation}
c_1 = \frac{4}{9} K_\Theta, ~~~ c_2 =  \frac{8}{9} K_\Theta, 
~~~ c_u = \frac{7}{9} K_u + \frac{8}{9} K_\Theta, ~~~ c_d = \frac{7}{9} K_d + \frac{8}{9} K_\Theta, ~~~ c_s = \frac{7}{9} K_s + \frac{8}{9} K_\Theta. ~~~~~
\end{equation}
In the case of the SM Higgs, we have $c_1 = 4/9$, $c_2 = 8/9$, $c_u = c_d = c_s = 5/3$~\cite{Gunion:1989we}, while for the up-philic scalar %of Sec.~\ref{sec:framework}
we have $c_u = g_u v /m_u$, $c_1 = c_2 = c_d = c_s = 0$.

We can use Eq.~(\ref{eq:L5}) to obtain the low-momentum couplings of the scalar $S$ to the Goldstone bosons. 
To obtain the physical couplings we must also account for $\eta-\eta'$ mixing. 
We diagonalize the system via the rotation
\begin{equation}
\left( 
\begin{array}{c}
\eta_8 \\
\eta_0
\end{array}
\right) = 
\left( 
\begin{array}{cc}
\cos\theta & \sin\theta \\
-\sin\theta & \cos\theta
\end{array}
\right) \left( 
\begin{array}{c}
\eta \\
\eta'
\end{array}
\right).
\label{eq:eta-rotate}
\end{equation}
The mixing angle has been studied on numerous occasions, see Ref.~\cite{Bickert:2016fgy} for a recent analysis and comparison with past results. 
We will take as a benchmark $\theta = -20^\circ$ in this work. For a scalar with arbitrary couplings at the quark level, then, Eqs.~\ref{eq:L5} and \ref{eq:eta-rotate} may be used to write its low energy interactions with mesons. There are both derivative interactions, from the first term of Eq.~(\ref{eq:L5}), and non-derivative couplings, from the second and third terms.

\subsection{Decay of scalar to pseudo-scalar mesons}

Here we compute the decay of the scalar to QCD mesons, augmenting the leading chiral Lagrangian treatment with experimentally extracted form factors for select modes which capture resonant effects that become important for $m_S$ above a few hundreds of MeV. 
%Here we compute the decay of the scalar to pions.
%We start from the Lagrangian (\ref{eq:L3}), rewritten here as 
%\begin{eqnarray}
%{\cal L} &= & - \frac{S}{v} \bigg\{ \frac{2}{9} K_\Theta \, \Theta^\mu_\mu + \frac{7}{9}\left(\frac{K_u+K_d}{2}\right) \, [m_u  \bar u u + m_d  \bar d d ]~~~~~~~~~~~ \nonumber \\
%&& ~~~~~~~~~~~~~+ \frac{7}{9} \left(\frac{K_u-K_d}{2}\right) \, [m_u  \bar u u - m_d  \bar d d ] 
% + \frac{7}{9}K_s \, m_s \bar s s  \bigg\}. 
%\label{eq:L3}
%\end{eqnarray}
It is convenient to define the form factors
\begin{eqnarray}
\langle \pi^i \pi^j | \,\Theta^\mu_\mu \, | 0\rangle = \Theta_\pi(s) \delta^{ij},  \nonumber \\
\langle \pi^i \pi^j |\, m_u \bar u u+ m_d \bar d d \, | 0\rangle = \Gamma_\pi(s) \delta^{ij},  \nonumber \\
\langle \pi^i \pi^j |\, m_u \bar u u- m_d \bar d d  \, | 0\rangle = \Omega_\pi(s) \delta^{ij}, \nonumber \\
\langle \pi^i \pi^j | \,m_s \bar s s \, | 0\rangle = \Delta_\pi(s) \delta^{ij}, 
\label{eq:pionFF}
\end{eqnarray}
with which we obtain the partial width for $S \rightarrow \pi \pi $:
\begin{equation}
\Gamma_{S\rightarrow \pi \pi} = \frac{3 G_F}{16 \sqrt{2} \pi m_S}  |G_\pi(m_S^2)|^2 \sqrt{1 -   \frac{4m_{\pi}^2}{m_S^2} },
\end{equation}
where 
\begin{equation}
G_\pi(s) = \frac{2}{9}\, K_\Theta \, \Theta_\pi + 
 \frac{7}{9}\left(\frac{K_u+K_d}{2}\right) \Gamma_\pi+
 \frac{7}{9}\left(\frac{K_u-K_d}{2}\right) \Omega_\pi +
  \frac{7}{9}K_s \Delta_\pi.
  \label{eq:pionG}
\end{equation}
In the low momentum limit, the form factors above can be evaluated through the use of the chiral Lagrangian using Eq.~(\ref{eq:matching}). We find 
\begin{equation}
\Theta_\pi(s) = s+2 m_\pi^2 , ~~~~ \Gamma_\pi(s) = m_\pi^2, ~~~~ \Omega_\pi(s) = 0, ~~~~ \Delta_\pi(s) = 0. 
\end{equation}
%We note that this agrees with the SM Higgs result in Refs.~\cite{Donoghue:1990xh,Winkler:2018qyg} for $K_\Theta = K_u = K_d = K_s = 1$. However, it seems to disagree with the general scalar coupling results considered in Ref.~\cite{Donoghue:1990xh} (their Eqs. 9, 10).  We also note the presence of the new form factor $\Omega_\pi(s)$ required in the case $K_u \neq K_d$.
We note that this agrees with the SM Higgs result in Refs.~\cite{Donoghue:1990xh,Winkler:2018qyg} for $K_\Theta = K_u = K_d = K_s = 1$, but note the presence of the new form factor $\Omega_\pi(s)$ required in the case $K_u \neq K_d$.

We can also compute the the partial widths for other final states. In particular, for $S \rightarrow K^+ K^-$ and $S \rightarrow K^0 \bar K^0$ we find
\begin{eqnarray}
\Gamma_{S\rightarrow K^+ K^-} & = & \frac{G_F}{8\sqrt{2} \pi m_S} |G_{K^+}(m_S^2)|^2 \sqrt{1 -   \frac{4m_{K}^2}{m_S^2} }, \nonumber \\
\Gamma_{S\rightarrow K^0 \bar K^0} & = & \frac{G_F}{8 \sqrt{2} \pi m_S}|G_{K^0}(m_S^2)|^2 \sqrt{1 -   \frac{4m_{K}^2}{m_S^2} },
\end{eqnarray}
where the form factors are defined analogously to those of the pions, Equations~(\ref{eq:pionFF},\ref{eq:pionG}). Using Eq.~(\ref{eq:matching}) we obtain in the low momentum limit 
\begin{eqnarray}
 \Theta_{K^+}(s) = \Theta_{K^0}(s) = s+2 m_K^2 , & ~~~~~~~~~~&  \Gamma_{K^+}(s) =  \Gamma_{K^0}(s) = \frac{1}{2}m_\pi^2,    \\
 \Omega_{K^+}(s) = - \Omega_{K^0}(s) = \frac{1}{2}m_\pi^2,  & ~~~~~~~~~~& \Delta_{K^+}(s) = \Delta_{K^0}(s) = m_K^2 -\frac{1}{2} m_\pi^2 .   \nonumber
\end{eqnarray}
For $\pi \pi$ and $KK$ final states the form factors at intermediate momentum scales of order 1 GeV have been extracted from data, and there are recent determinations~\cite{Winkler:2018qyg} for
$\Theta_{\pi, K},\Gamma_{\pi,K}, \Delta_{\pi,K}$.  No results are available for the form factors $\Omega_{\pi,K}$, and we will therefore simply use the Chiral Lagrangian results given above. 
Similarly, for final states involving $\eta, \eta'$ determinations of the corresponding form factors  from data are not available. At the higher scalar masses where decays to heavier mesons become important, we expect in any case that our chiral Lagrangian approach can only provide a rough estimate of the interactions of $S$.
%In light of this, we will simply compute the partial widths from the chiral Lagrangian in Equations (\ref{eq:L6},\ref{eq:L8}) above. The partial decay widths for the $\pi^0 \eta$, $\pi^0 \eta'$, and $\eta \eta$ channels are 
%In light of this, we will simply compute the partial decay width for the $\pi^0 \eta$ channel, and treat the heavier decay channels together below. From the chiral Lagrangian above, we have
%\begin{eqnarray}
%\Gamma_{S\rightarrow \pi^0 \eta} & = &  \frac{a_{S\pi^0 \eta}^2}{16 \pi m_S} \lambda^{1/2}\left(1, \frac{m_\pi^2}{m_S^2},  \frac{m_\eta^2}{m_S^2} \right) \\
%\Gamma_{S\rightarrow \pi^0 \eta'} & = &  \frac{a_{S\pi^0 \eta'}^2}{16 \pi m_S} \lambda^{1/2}\left(1, \frac{m_\pi^2}{m_S^2},  \frac{m_{\eta'}^2}{m_S^2} \right), \nonumber \\
%\Gamma_{S\rightarrow  \eta \eta} & = &  \frac{1}{32 \pi m_S} \left[\frac{c_1}{2 v} (m_S^2-2 m_\eta^2) + a_{S\eta\eta}^2 \right]^2 \sqrt{1 -   \frac{4m_{\eta}^2}{m_S^2} }, \nonumber \\
%\Gamma_{S\rightarrow \eta \eta'} & = &  \frac{a_{S\eta\eta'}^2}{16 \pi m_S} \lambda^{1/2}\left(1, \frac{m_\eta^2}{m_S^2},  \frac{m_{\eta'}^2}{m_S^2} \right).  
%\end{eqnarray}
%with
%\begin{equation}
%a_{S\pi^0\eta} = \frac{1}{v}  (c_u - c_d) m_\pi^2 \left( \frac{c_\theta}{2\sqrt{3}}  - \frac{s_\theta}{\sqrt{6}} \right) .
%\end{equation}

%\AI{Moved from framework section, start from here when discussing matching to partonic calculation.}
In light of this, to calculate the decay width of the $S$ at low mass we start with the results above for the $\pi \pi$ and $K K$ modes, replacing the leading chiral Lagrangian form factors with those experimentally extracted from data where available~\cite{Winkler:2018qyg}. In addition to these decays, we account for other modes including $4\pi$, $\pi \eta(')$ and $\eta(')\eta(')$ with an additional channel
\be
\Gamma_{S \to~\mathrm{other\ hadrons}} \sim m_S^3 \sqrt{1 - 16 m_\pi^2 / m_S^2}.
\ee
The overall normalization of this contribution is fixed by requiring that the total $S$ decay width to mesons matches the partonic width of $S$
%$S \to q \bar{q}$ width
at $m_S = 2~\mathrm{GeV}$.
%, which in general is given by \AI{wrote q instead of u},
%\begin{equation}
%\Gamma_{S} = \sum_{m_S > 2 m_q} {\frac{3 \kappa_q^2 m_q^2 m_S}{8 \pi v^2}} + \frac{\alpha_s^2 \kappa_G^2 m_S^3}{72 \pi^3 v^2} .
%\end{equation}
Above $m_S = 2~\mathrm{GeV}$, we simply use the partonic width.

\subsection{Scalar coupling to nucleons}

%Here we discuss the effective coupling of the scalar mediator to nucleons. We start from the Lagrangian (\ref{eq:L3}) and the definition of the trace of the energy momentum tensor (\ref{eq:trEM}). Taking the matrix element of the energy momentum tensor between nucleons at zero momentum, we have 
%\begin{eqnarray}
%m_N  = \langle N | \, \Theta_\mu^\mu \, |N \rangle  & = &  \bigg\langle N \bigg| \!-\!\frac{9\alpha_s}{8 \pi} G_{\mu\nu}^a G^{\mu\nu a}  \, \bigg| N \bigg\rangle 
%+\sum_{q = u,d,s} \langle N | \, m_q \bar q q \, |N \rangle \nonumber  \\
%& = & m_N f_{TG}^{(N)} + \sum_{q = u,d,s} m_N f_{Tq}^{(N)}.
%\end{eqnarray}
%This defines  the gluon matrix element $f_{TG}^{(N)}$ in terms of those of the light quark $f_{Tq}^{(N)}$,
%\begin{equation}
%f_{TG}^{(N)} =1- \sum_{q=u,d,s} f_{Tq}^{(N)}
%\end{equation}
Several calculations in this work have relied on the the effective coupling of $S$ to nucleons. This is obtained by matching Eq.~(\ref{eq:L3}) to the nucleon level interaction~\cite{Shifman:1978zn},
\begin{equation}
{\cal L} \supset - y_{SNN} S \bar N N.
\label{eq:ySNN}
\end{equation}
where $N = p, n$. We find 
\begin{eqnarray}
y_{SNN} & = &  \frac{1}{v}\left[   \frac{2}{9} \, K_\Theta \,  \langle N | \, \Theta_\mu^\mu \, |N \rangle + \frac{7}{9}  \sum_{q = u,d,s} \, K_q  \,\langle N | \, m_q \bar q q \, |N \rangle   \right], \nonumber \\
& = &   \frac{m_N}{v}\left[   \frac{2}{9} \,  K_\Theta + \frac{7}{9}  \sum_{q = u,d,s} K_q \, f_{Tq}^{(N)}  \right],
\end{eqnarray}
where we have used $\langle N | \, \Theta_\mu^\mu \, |N \rangle = m_N$ and $\langle N | \, m_q \bar q q \, |N \rangle = f_{Tq}^{(N)}$ for the nucleon matrix elements. For the case of the up-philic scalar we have $y_{SNN} = g_u f_{Tu}^{(N)}  m_N/m_u$.

%\end{thebibliography}

%\end{document}

\bibliography{up-DM}

\end{document}